%% file: wind_drifter_paper_v3.tex
\documentclass[twocol]{ametsocV6.1}

\makeatletter
\renewcommand{\p@subsection}{\thesection}
\makeatother

\title{Wind vector retrieval from miniaturized wave-enabled sea-surface drifters}

\authors{Alexey Mironov\correspondingauthor{Alexey Mironov, alexey.mironov@eodyn.com}\aff{a}}

\affiliation{\aff{a}{eOdyn, Plouzané, France}}

\abstract{Open-ocean in situ winds are sparse, limiting the calibration and
validation of satellite wind products. We evaluate wind-vector retrieval
from MELODI (Miniaturized Electronics Lagrangian Oceanographic DrIfter), a
compact freely drifting GNSS/IMU wave buoy. Wind speed is inferred from
the shape of the measured wave acceleration spectrum. A regularized
regression with a gated nonlinear correction (Wind Inversion using
Tikhonov Regularization, WITR), trained on scatterometer winds and
evaluated by leave-one-buoy-out cross-validation against ASCAT MetOp-B/C
and HY-2B/C, gives a root-mean-square difference (RMSD) of
0.91~m~s$^{-1}$. Symbolic regression distills the trained model into
compact spectrum-only and motion-enhanced equations suitable for onboard
use; the reduced drag law gives 0.93--0.94~m~s$^{-1}$. Wind direction is
estimated, without any fitted reference-wind regression, from
IMU-derived directional wave moments in the 0.60--0.90-Hz wind-sea band, with a circular mean absolute angular difference
(MAAD) of 9.4$^{\circ}$ against ASCAT. These differences are comparable to
published scatterometer--moored-buoy comparisons and improve
substantially on the traditional single-band Toba inversion. Validation
is strongest from 4 to 14~m~s$^{-1}$ and sparse above approximately
16~m~s$^{-1}$; stronger winds are assigned low confidence. The results
show that small wave-sensing drifters, calibrated to the well-established
scatterometer wind scale, can provide useful wind-vector observations for
validating satellite wind products and for atmospheric and wave modeling;
stronger-wind sampling remains a priority.}

\begin{document}

\maketitle

\statement
Satellite sensors measure winds over most of the ocean every day, but the
globally representative in situ reference winds needed for their absolute
calibration remain sparse, and calibrating and
validating satellite wind sensors depends on surface observations matched in
space and time to each overpass---precisely what is missing over the open
ocean. Compact drifting buoys already estimate the open-ocean wind vector from
the waves they ride, but not very accurately; we show that a small, low-cost
GNSS/IMU drifter can do markedly better---sharpening both wind speed (from the
wave spectrum) and direction (from the buoy's own motion) to an accuracy
consistent with scatterometer winds and comparable to published
scatterometer--buoy differences, close to a factor-of-two improvement
over earlier single-band drifter methods. Because the governing relationships are
simple enough to run onboard, fleets of these biodegradable drifters could
affordably fill the open-ocean wind-observation gap.

\section{Introduction}

Knowledge of the near-surface wind vector over the ocean is essential for
weather forecasting, climate monitoring, air--sea flux estimation, and
maritime operations. Over most of the world ocean this wind is measured
routinely only from space. The virtual scatterometer
constellation---the Advanced Scatterometer (ASCAT) on MetOp-B and
MetOp-C, the Haiyang-2 series (HY-2B/C), and OceanSat-3---retrieves the
surface wind vector from microwave backscatter with accuracies of order
1~m~s$^{-1}$ in speed and 15--20$^{\circ}$ in direction
\citep{Stoffelen1998}, revisiting a given location two to four times per
day; its members, though distinct instruments, are intercalibrated to
good accuracy \citep{WangEA2017,WangEA2021,ZhaoEA2023,VogelzangStoffelen2022}.
Synthetic aperture radars provide high-resolution wind fields but with
limited temporal sampling and lower wind-vector accuracy
\citep{Mouche2019,ChenEA2025}; altimeters deliver along-track wind speed
estimates \citep{Young1999,Zieger2009,RibalYoung2019}; and the SWOT
mission needs the local wind-wave field to correct its sea-state bias
\citep{Morrow2019}. None of these sensors---nor the passive radiometers
that complement them, nor the atmospheric and wave models they
feed---carries an absolute calibration: turning this coverage into
trustworthy measurements requires a globally representative network of
well-calibrated in situ winds, collocated within about half an hour of an
overpass \citep{Stoffelen2015}. The existing network of moored
meteorological buoys---approximately 400 stations operated by the
National Data Buoy Center (NDBC), M\'{e}t\'{e}o-France, the tropical
moored arrays (TAO/TRITON, PIRATA, RAMA), and partner agencies---covers
less than 1\% of the ocean surface, concentrated along coastlines and in
the tropics \citep{KentEA2019,VillasBoas2019}; vast regions including
the Southern Ocean, remote Pacific, and much of the Indian Ocean lack the
sustained in situ winds needed for calibration and validation
\citep{Bourassa2019}.

Surface drifters offer a potential solution. The Global Drifter Program
(GDP) maintains approximately 1\,500 active drifters measuring sea surface
temperature and near-surface currents \citep{Centurioni2018}. These
platforms are inexpensive, globally distributed, and already demonstrate
99\% data return rates. If drifters could also measure wind, they would
provide the missing open-ocean in situ observations at a fraction of the
cost of moored buoys.

The physical basis for this is the equilibrium-range scaling of
wind-generated gravity waves \citep{Toba1973,Phillips1985,Donelan1985},
reviewed in section~\ref{sec:theory}: the spectral level in the wind-sea
range is set by the wind friction velocity, so the wind can in principle
be recovered by inverting the measured wave spectrum.

Recent advances in miniaturized wave-sensing drifters have made this
practical. \citet{Herbers2012} showed that drifters tracked by the Global Positioning System (GPS) reliably
measure directional wave spectra; \citet{Thomson2012,Thomson2024} developed
the SWIFT (Surface Wave Instrument Float with Tracking) and microSWIFT platforms; and
\citet{YurovskyDulov2020,YurovskyKudinov2026} used buoys based on micro-electro-mechanical systems (MEMS) to
resolve the short wind-wave spectrum from a very small platform. Building on
this, several studies have retrieved wind from such spectra:
\citet{Voermans2020} obtained $\sim$2~m~s$^{-1}$ speed RMSD from the Sofar
Spotter (40~cm diameter), \citet{Dorsay2023} reached $\sim$1~m~s$^{-1}$
across a global Spotter fleet (physics-based, spectral-balance, and neural
approaches), \citet{Shimura2022} confirmed the method for small GPS buoys,
and \citet{Jiang2022} trained neural networks on NDBC directional
coefficients (1.1~m~s$^{-1}$ speed, 14$^{\circ}$ direction above
7~m~s$^{-1}$).

Despite this progress, existing drifter retrievals share important
limitations. The traditional approach applies the Toba relationship to a
single spectral band, ignoring that different frequencies respond to wind
changes on different timescales and that spectral shape depends on wave age,
atmospheric stability, and swell contamination. Published
single-band retrievals accordingly show speed RMSD $\geq$2~m~s$^{-1}$ and
unreliable direction, especially at low wind: \citet{Voermans2020} reported
$\sim$20$^{\circ}$ accuracy only above 10~m~s$^{-1}$, and the yearlong
evaluation of \citet{BeckmanLong2022} found a Spotter direction RMSD of
$95.7^{\circ}$ ($R^2=0.32$). Our aim is therefore not to establish that
drifters can sense the wind vector---by now well demonstrated, most
prominently by the Sofar Spotter network \citep{Dorsay2023}---but to improve
the accuracy of both components, especially the low-wind direction, on a
smaller platform, and to evaluate the result against multi-mission
scatterometer winds.

This paper presents a wind vector retrieval method for the MELODI
(Miniaturized Electronics Lagrangian Oceanographic DrIfter) buoy
\citep{MironovCharron2023,MironovCharron2024}, a compact freely drifting
platform (24-cm hull diameter, smaller than the 40--42-cm Spotter) that
measures the surface wave field with a multi-constellation Global
Navigation Satellite System (GNSS) receiver and a 9-axis inertial
measurement unit (IMU). Our
contribution is fourfold. First, we retrieve wind speed with a multi-band
spectral inversion---the Wind Inversion using Tikhonov Regularization
\citep[WITR;][]{Hoerl1970}, a regularized linear inversion with a gated
gradient-boosted residual correction---that exploits the full shape of the
measured acceleration spectrum rather than a single equilibrium-range level,
and we distill the
trained model into compact, interpretable analytical expressions through
symbolic regression \citep[PySR;][]{Cranmer2023}, recovering interpretable
wind--wave structure including wave-age modulation, smooth-to-rough
transition behaviour, and high-wind drag reduction. Second, we retrieve wind
direction directly from
directional wave moments derived from the IMU, sharpening the component where
existing drifter retrievals are weakest, especially at low wind. Third, we evaluate
both components against multi-mission scatterometer winds: by
leave-one-buoy-out cross-validation for the scatterometer-trained speed
model, and without any fitted reference-wind regression for the
direction retrieval. Fourth, to show
that the retrievals can be applied at fleet scale, we demonstrate how the
speed and direction estimates are assembled and quality-controlled into a
wind-vector dataset spanning a multi-buoy fleet across four ocean basins.

The method is developed and evaluated on three complementary datasets of
increasing sensor fidelity---satellite-transmitted acceleration spectra,
recovered high-rate IMU records, and a dedicated GNSS+IMU field
deployment---trained against scatterometer winds, with ERA5 reanalysis
used during feature development.

\section{Theoretical background and motivation}\label{sec:theory}

The physical basis for wind estimation from wave observations is well
established. \citet{Toba1973} showed that the spectral energy density in
the equilibrium range of wind-generated gravity waves scales with the
wind friction velocity as $S(f) \propto g\,u_*\,f^{-4}$.
\citet{Phillips1985} and \citet{Kitaigorodskii1983} provided complementary
theoretical frameworks linking the equilibrium spectral level to wind
forcing through energy cascades and wave breaking; \citet{Donelan1985} and
\citet{Thomson2013} confirmed the relationship with field data, the latter
documenting the equilibrium range and its friction-velocity scaling directly
at Ocean Weather Station~P. From a measured wave spectrum the wind can
therefore, in principle, be recovered by inverting this scaling.

\begin{figure}[t]
  \noindent\includegraphics[width=19pc]{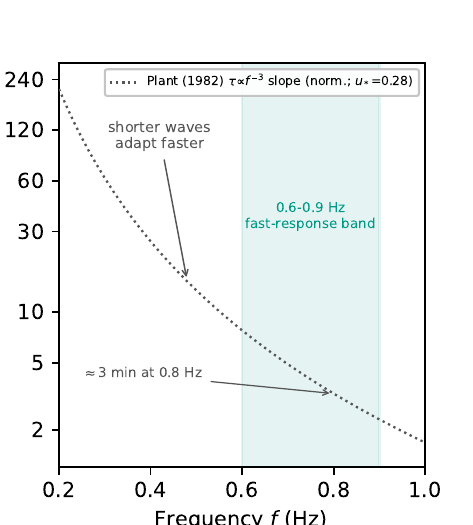}
  \caption{Spectral adaptation timescale $\tau$ as a function of frequency. The \citet{Plant1982} wind-input growth rate gives $\tau \sim 1/\gamma \propto f^{-3}$, with the response time set to several hours near the 0.2~Hz spectral peak: shorter, higher-frequency waves equilibrate within minutes, whereas the energetic low-frequency bands lag the wind by hours. This motivates restricting the high-resolution retrieval to the high-frequency bands. Only the theoretical scaling is shown; the lag resolvable from reanalysis collocation ($\gtrsim$1~h, set by the hourly ERA5 cadence) is too coarse to measure the fast-band timescales directly.}
  \label{fig:adapt}
\end{figure}

Formally, in the equilibrium range the displacement spectral density
follows the Toba scaling:
\begin{equation}\label{eq:toba}
S_\eta(f) = \alpha\,g\,u_*\,(2\pi)^{-3}\,f^{-4} \,,
\end{equation}
where $g$ is gravitational acceleration, $u_*$ is the wind friction
velocity, and $\alpha = 0.062$ is the equilibrium constant. This
relationship means that the spectral level in the equilibrium range is
directly proportional to the wind friction velocity, providing the
physical basis for wind speed retrieval. Equilibration is strongly
frequency-dependent: with the \citet{Plant1982} wind-input growth rate
($\gamma \propto (u_*/c)^2 f$, phase speed $c = g/2\pi f$), the
input timescale $\tau \sim 1/\gamma \propto f^{-3}$ falls sharply with
frequency, and with the response time set to several hours near the
spectral peak ($\sim$0.2~Hz) this scaling predicts equilibration within
$\sim$2--8~min in the 0.6--0.9~Hz wind-sea band but multi-hour lags at
low frequency (Fig.~\ref{fig:adapt}). These sub-hour timescales cannot be
confirmed against reanalysis---the hourly ERA5 cadence and $\sim$22-min
session spacing floor the resolvable lag near one hour---but they are the
physical basis for restricting the high-resolution retrieval to the
high-frequency bands, where the spectral level is an instantaneous wind
proxy.

The Toba spectral level parameter is computed for each of four frequency
bands---low (LO, 0.12--0.30~Hz), mid (MID, 0.25--0.50~Hz), high (HI,
0.45--0.75~Hz), and very high (VHI, 0.70--1.00~Hz):
\begin{equation}\label{eq:beta4}
\beta_4 = \mathrm{median}\!\left[S_\eta(f)\,f^4\right]_{f \in \mathrm{band}} \,,
\end{equation}
and the corresponding 10-m wind speed estimate $U_{10}^{\mathrm{Toba}}$
is obtained by inverting Eq.~(\ref{eq:toba}). Because the band level
$\beta_4 = \mathrm{median}[S_\eta(f)\,f^4] = \alpha\,g\,u_*\,(2\pi)^{-3}$
is linear in $u_*$, the friction velocity follows directly from the
spectral level:
\begin{equation}\label{eq:toba_inversion}
u_* = \frac{\beta_4\,(2\pi)^3}{\alpha\,g} \,.
\end{equation}
Converting $u_*$ to $U_{10}$ uses the wind-dependent neutral drag
coefficient of \citet{LargePond1981}, $C_D = (0.49 + 0.065\,U_{10})\times
10^{-3}$ with $U_{10} = u_*/\sqrt{C_D}$, solved by fixed-point iteration
(3--5 steps).

Equation~(\ref{eq:toba_inversion}) underlies the established reference
method operational on the Sofar Spotter network
\citep{Voermans2020,Dorsay2023}, which locates the equilibrium range,
extracts $u_*$ from that single level, and takes direction from the
band-averaged first-order directional moments of the wave spectrum
(section~\ref{sec:direction}). It performs well at moderate winds (RMSD
$\approx 2$~m~s$^{-1}$ over 3--12~m~s$^{-1}$; \citealp{Voermans2020}) but,
reading a single level, inherits two limitations. First, the $f^{-4}$ shape
is not uniquely dynamical---\citet{belcher1997breaking} show it also arises
kinematically from breaking-crest geometry---so a fixed equilibrium constant
$\alpha$ cannot be optimal across sea states. Second, the level-to-$u_*$
mapping saturates at high winds \citep{Davis2023}. Rather than commit to one
band and a fixed $\alpha$, we retain the four-band levels of
Eq.~(\ref{eq:beta4}) as inputs and let the regression of
section~\ref{sec:tr} learn their wind dependence from collocated reference
winds, absorbing the wave-age, stability, and breaking-amplitude effects the
analytic inversion ignores.

\section{MELODI drifter system}\label{sec:system}

\subsection{Design and sensor payload}

\begin{figure}[t]
\centerline{%
  \raisebox{0.75cm}{\includegraphics[height=3cm]{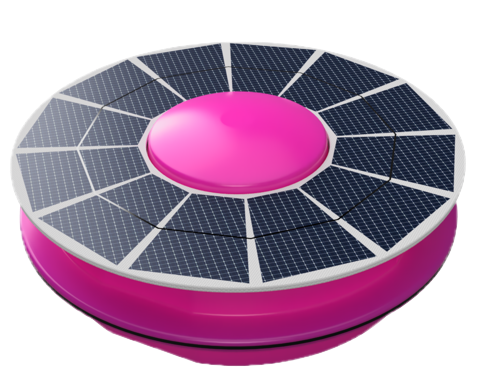}}\quad
  \includegraphics[height=4.5cm]{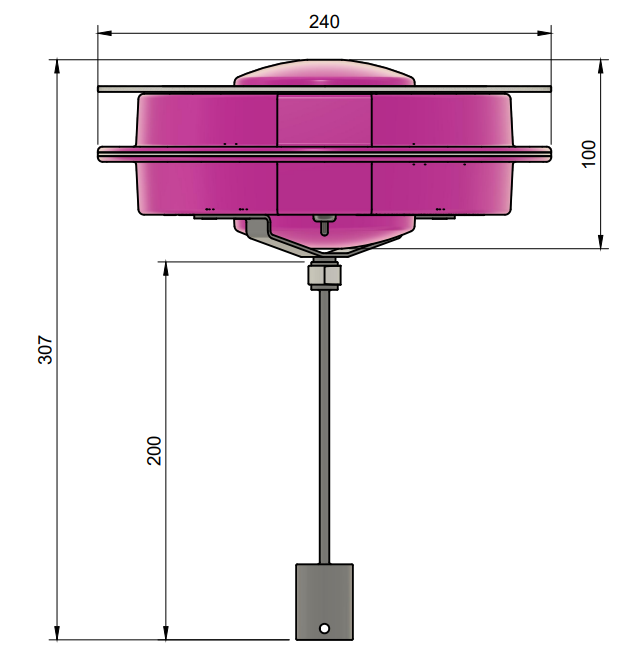}%
}
  \caption{The MELODI surface drifter: (left) exterior view showing the
  solar-panel array and biodegradable polybutylene succinate (PBS) hull;
  (right) cross-section with
  principal dimensions in millimetres (240~mm hull diameter, 307~mm overall
  height, 100~mm disc spacing, 200~mm stabilizing pole, and the ballast
  weight that maintains vertical orientation). Sensor specifications are
  listed in Table~\ref{tab:sensors}.}
  \label{fig:dims}
\end{figure}

MELODI is a low-cost, expendable surface drifter developed by
eOdyn (Plouzan\'{e}, France) for operational sea-state monitoring
(Fig.~\ref{fig:dims}). The hull is formed by two symmetrical
half-shells, 240~mm in diameter and 100~mm in height, with a total
mass of approximately 1.3~kg including a 200~mm stabilizing pole and
ballast weight (307~mm overall height; Fig.~\ref{fig:dims}).
It is moulded from polybutylene succinate (PBS), a bio-sourced
bioplastic that degrades into water and CO$_2$ through natural
enzymatic processes, avoiding persistent microplastic release
\citep{XuGuo2010,Aliotta2022,MironovCharron2024}. The low underwater profile ($\sim$100~mm)
minimizes direct wind forcing on the hull; field estimates put the
wind-induced hull drift near 1\% of the 10-m wind speed, with individual
estimates ranging from 0.7\% to 1.4\%.

The sensor payload and onboard electronics (Table~\ref{tab:sensors})
comprise a multi-constellation GNSS receiver (2.5-m fix accuracy; sampled
at 4~Hz on Tier-C hardware, yielding the horizontal velocities from which
wave spectra are derived); a 9-axis IMU (accelerometer, gyroscope,
magnetometer), acquired at 3.2~Hz and processed onboard into an attitude
and heading reference system (AHRS) output for the recovered fleet
(Tier~B) or logged raw at 100~Hz on the newest hardware (Tier~C); a sea
surface temperature sensor; a 6-W solar array with maximum-power-point
tracking charging 14\,000~mAh of lithium-ion cells, so that deployment
duration is set by fouling, damage, or loss rather than by energy; an
Iridium Short Burst Data (SBD) modem providing bidirectional global
communication (99.8\% data return) and remote reconfiguration; and a
micro-SD card buffering the full-rate records retrieved from recovered
units.

The onboard microcontroller reconstructs
vertical acceleration from the raw IMU via a Kalman filter and computes the
omnidirectional acceleration spectrum $S_{\mathrm{acc}}(f)$ over $\sim$22-min
windows (0.02--1~Hz, 128~bins), converted to displacement spectral density
via
\begin{equation}\label{eq:acc2disp}
S_\eta(f) = \frac{S_{\mathrm{acc}}(f)}{(2\pi f)^4} \,,
\end{equation}
from which $H_s = 4\sqrt{m_0}$ ($m_0 = \int S_\eta\,\mathrm{d}f$). Surface
currents are estimated from successive GNSS fixes (600-s finite differences,
median-filtered). Section~\ref{sec:processing} describes the analysis-side
processing applied to both onboard and recovered spectra.

\begin{table*}[t]
\caption{MELODI onboard sensor specifications.}\label{tab:sensors}
\begin{center}
\begin{tabular}{lcccc}
\hline\hline
Sensor & Range & Precision & Rate & Units \\
\hline
GNSS receiver   & ---           & 2.5~m     & 4~Hz$^*$    & m \\
Accelerometer   & $\pm 4$       & 0.061     & 3.2--100~Hz & g ($\times 10^{-3}$) \\
Gyroscope       & $\pm 250$     & 4.375     & 3.2--100~Hz & $^{\circ}$s$^{-1}$ ($\times 10^{-3}$)\\
Magnetometer    & $\pm 800$     & 0.02      & 3.2~Hz      & $\mu$T \\
Temperature     & $-$10 to $+$85 & 0.0625   & 0.03~Hz     & $^{\circ}$C \\
\hline\hline
\end{tabular}
\end{center}
{\footnotesize $^*$4~Hz for Tier~C hardware; Tier~B buoys use onboard
  AHRS at 3.2~Hz. Precision values are per least significant bit.}
\end{table*}

\subsection{Data transmission and recovery}

MELODI operates in two data modes that define the information content
available for wind retrieval:

\textit{Satellite-transmitted mode (Tier~A).} Iridium SBD messages are
emitted every 30~minutes and carry three consecutive GNSS positions
(10-min effective resolution), one SST value, and the significant wave
height $H_s$. The full 128-bin acceleration spectrum
$S_{\mathrm{acc}}(f)$ is transmitted separately on a configurable
cadence of one or three hours, compressed to fit within the SBD
payload. No raw IMU data are available in this mode; consequently,
only omnidirectional spectral features can be extracted for wind speed
estimation. This is the standard operational mode for the deployed
fleet used in this study.

\textit{Recovered data mode (Tier~B and C).} When buoys are physically
recovered, the onboard micro-SD card provides full-rate sensor data. For
Tier~B buoys, this includes AHRS output (vertical acceleration,
pitch, roll, heading) at 3.2~Hz and raw 9-axis IMU data at the same rate.
For Tier~C buoys (newest hardware), GNSS velocity is logged at 4~Hz and
raw IMU at 100~Hz. Each acquisition session spans approximately
22~minutes ($\sim$4\,200 samples at 3.2~Hz) and is recorded to onboard
memory every 30~minutes, so a recovered buoy provides a continuous record at
half-hourly cadence (the satellite link transmits a subset at the
configurable 1- or 3-hour cadence). The high-rate data enables extraction of
directional wave moments from the IMU gyroscope and magnetometer channels,
which are
essential for wind direction retrieval.

\section{Datasets}\label{sec:datasets}

The experimental dataset is organized into three tiers defined by the
available sensor data, supplemented by two independent reference sources
(Table~\ref{tab:datasets}). This section describes the data; the
procedure used to collocate the reference products with the MELODI
sessions is given in section~\ref{sec:collocation}.

\subsection{Satellite-transmitted spectra (Tier~A)}

Twenty-four MELODI buoys deployed during the 2024--2025 ocean
campaigns, with collocated sessions spanning March 2025 to March 2026 (section~\ref{sec:campaigns}) transmitted onboard-computed acceleration spectra
$S_{\mathrm{acc}}(f)$ via Iridium SBD, yielding approximately 8\,000
sessions over deployment periods of 2--7 months per buoy. Each session
provides a 128-bin acceleration spectrum in the range 0.035--1.0~Hz, from
which omnidirectional spectral features (band levels, spectral
slopes, Toba wind estimates) can be computed. No raw IMU data are
available; therefore, Tier~A supports wind speed retrieval only. These
buoys span the northeast Atlantic, Norwegian Sea, Mediterranean, and
tropical Atlantic, providing the broadest geographic coverage and
demonstrating that the wind retrieval method generalizes beyond the
Tier~B training set.

\subsection{Recovered IMU data (Tier~B)}\label{sec:campaigns}

Four buoys were physically recovered after their deployments, providing
full-rate sensor data from the onboard micro-SD storage. Three buoys
were deployed during the ESA Ocean Training Courses 2025 (OTC25) campaign
from the tall ship \textit{Statsraad Lehmkuhl}, and one during the
EXPLOI campaign (Exp\'{e}dition Plastique Oc\'{e}an Indien; COI, AFD,
FFEM; scientific partner: IRD). These four buoys constitute the primary
dataset for model development and validation:
\begin{itemize}
\item \textit{OTC25\_04} (Bay of Biscay, 47--49$^{\circ}$N, May--November
  2025): 9\,129 sessions over 6 months, capturing the full seasonal
  cycle from spring calms to autumn gales (wind speeds 0--21~m~s$^{-1}$).
\item \textit{EXPLOI\_06} (Indian Ocean, 18--22$^{\circ}$S, June--December
  2025): 8\,603 sessions over 6 months in the trade wind belt, providing
  the most stable conditions (mean wind 7.8~m~s$^{-1}$) and the lowest
  validation errors.
\item \textit{OTC25\_16} (northeast Atlantic, $\sim$54$^{\circ}$N, deployed at
  54.9$^{\circ}$N, 17.0$^{\circ}$W and drifting east to 54.1$^{\circ}$N,
  10.0$^{\circ}$W by 28~July 2025): 3\,384 sessions spanning 2 months,
  including several strong-wind events with sustained winds above 15~m~s$^{-1}$.
\item \textit{OTC25\_20} (Strait of Gibraltar, 35--36$^{\circ}$N, May--June
  2025): 626 sessions over 2 weeks in a region characterized by strong
  tidal currents and channeled winds.
\end{itemize}
Together these buoys provide 21\,742 sessions with AHRS data at 3.2~Hz
(vertical acceleration, pitch, roll, heading) and raw 9-axis IMU data,
enabling both wind speed and direction retrieval.

\subsection{GNSS+IMU test deployment (Tier~C)}

In March 2026, four buoys of the newest hardware revision (N1--N4) were
deployed in the Bay of Biscay ($\sim$48.3$^{\circ}$N, 4.6$^{\circ}$W) for
a 48-hour validation test. These buoys provide GNSS velocity at 4~Hz and
raw IMU at 100~Hz, yielding 101 sessions with enhanced directional wave
moments. The dominant first-order directional moment reaches
$|a_1| \approx 0.85$ in the wind-sea band, compared to $|a_1| \approx
0.65$ for Tier~B AHRS data, consistent with improved directional
resolution at the higher sampling rate. While the dataset is small, it demonstrates
the capability of the next generation of MELODI hardware.

\subsection{ERA5 reanalysis}\label{sec:era5}

All MELODI sessions are collocated with the ERA5 reanalysis
\citep{Hersbach2020} at its native 0.25$^{\circ}$, hourly resolution,
providing 10-m wind speed $U_{10}$, wind direction $\theta_w$, wind gusts
$U_g$, and 2-m air temperature $T_{\mathrm{air}}$. The resulting collocation dataset
comprises 21\,742 matched sessions (100\% yield) with ERA5 wind speeds
ranging from 0 to 19~m~s$^{-1}$ (mean 7.2~m~s$^{-1}$). An air--sea
temperature difference $\Delta T = T_{\mathrm{SST}} - T_{\mathrm{air}}$ is
computed for each session as a proxy for atmospheric stability, which
plays a role in the wind--wave coupling. ERA5 wind speed is used during
feature development and feature scoping for the wind speed retrieval model
(section~\ref{sec:speed}); the final model is trained on the
scatterometer winds described below and evaluated by leave-one-buoy-out
(LOBO) cross-validation (section~\ref{sec:tr_rect}).
ERA5 surface winds carry known limitations relevant here: the effective
resolution of the wind field is far coarser than the nominal
0.25$^{\circ}$ grid, mesoscale variability and transients are
under-represented, and persistent regional biases of up to
$\sim$0.5~m~s$^{-1}$ remain \citep{BelmonteStoffelen2019}. Training the
retrieval on scatterometer winds rather than on ERA5 avoids inheriting
these biases (scatterometer-corrected ERA5 products exist but would
introduce the scatterometer indirectly).

\subsection{Scatterometer winds}\label{sec:scat}

The retrieved winds are evaluated against Level-2 scatterometer products
from four satellite instruments---ASCAT on MetOp-B and MetOp-C (C-band,
12.5~km; \citealp{Stoffelen1998}) and the HY-2 scatterometers (HSCAT) on
HY-2B and HY-2C (Ku-band,
25~km)---obtained as swath-level Level-2B (L2B) data in the standard
format of the Royal Netherlands Meteorological Institute (KNMI)
(March 2025--March 2026; Table~\ref{tab:datasets}). These serve as the training target and
leave-one-buoy-out cross-validation reference for the speed model and as an
independent reference for the direction retrieval; the collocation procedure
(section~\ref{sec:collocation}) matches each session to the nearest
quality-screened cells within 25~km and 30~min. The 12.5- and 25-km
figures are grid spacings rather than true resolutions: the ASCAT 12.5-km
coastal product resolves $\approx$28~km \citep{Vogelzang2017}, and both
MetOp satellites contribute L2B products. The distributed products
retain no quality-controlled winds below 3~m~s$^{-1}$, which sets the
low-wind floor of all scatterometer comparisons in this study. A dedicated
re-collocation retaining the cells flagged only as \texttt{small\_wind}
(641 matches, 0.3--3~m~s$^{-1}$) shows the retrieval tracking these low
scatterometer winds with 1.0--1.1~m~s$^{-1}$ RMSD and bias below
$+$0.5~m~s$^{-1}$ down to $\approx$1~m~s$^{-1}$, the drifter's own
variability (1.2 versus 0.7~m~s$^{-1}$) dominating the spread.

The asymmetry has a physical origin: both systems sense the wind through its
wave manifestation, but at very different wavelengths---the scatterometer
through 1--5-cm capillary--gravity roughness, which responds almost
instantly to the lightest airflow, and MELODI through metre-scale gravity
waves ($\lambda \approx 2$--100~m across the exploited bands) whose
shortest resolved components travel at $c \approx 1.7$~m~s$^{-1}$ and
cease to be actively forced once the wind falls below
$\approx$2~m~s$^{-1}$.

\begin{table*}[t]
\caption{Summary of MELODI datasets and reference collocations. $N$
  denotes number of sessions (or collocations for reference sources).
  Wind speed range refers to the reference
  (ERA5 or scatterometer).}\label{tab:datasets}
\begin{center}
\begin{tabular}{llrrccc}
\hline\hline
Tier / Source & Buoys & $N$ & Period & Sensors & Sampling/grid & Wind (m~s$^{-1}$) \\
\hline
\multicolumn{7}{c}{\textit{MELODI buoy data}} \\
A (satellite spectra)    & 24 & 8\,033  & 2025--2026      & $S_{\mathrm{acc}}$ 128 bins & 30~min & --- \\
B (recovered IMU)        & 4  & 21\,742 & May--Dec 2025   & AHRS 3.2~Hz + raw IMU      & 22~min & --- \\
C (GNSS+IMU test)        & 4  & 101     & Mar 2026        & GNSS 4~Hz + IMU 100~Hz       & 22~min & --- \\
\hline
\multicolumn{7}{c}{\textit{Reference collocations}} \\
ERA5 reanalysis          & ---& 21\,742 & May--Dec 2025   & 0.25$^{\circ}$, hourly      & $\pm$30~min & 0--19 \\
ASCAT MetOp-B L2B        & 25 & 1\,424  & Mar 2025--Mar 2026 & C-band, 12.5~km          & $\pm$30~min & 3--20 \\
ASCAT MetOp-C L2B        & 25 & 1\,275  & Mar 2025--Mar 2026 & C-band, 12.5~km          & $\pm$30~min & 3--17 \\
HY-2B HSCAT L2B          & 26 & 1\,906  & Mar 2025--Mar 2026 & Ku-band, 25~km           & $\pm$30~min & 3--21 \\
HY-2C HSCAT L2B          & 25 & 2\,024  & Mar 2025--Mar 2026 & Ku-band, 25~km           & $\pm$30~min & 3--20 \\
\textbf{All L2B}         & \textbf{26} & \textbf{6\,629} & & & & \textbf{3--21} \\
\hline\hline
\end{tabular}
\end{center}
{\footnotesize The four L2B instruments above form the full collocation
set. The tight quality-controlled subset used for cross-validation
additionally draws on the ASCAT MetOp-B OSI-104 coastal product, giving five
scatterometer products across the same four satellites.}
\end{table*}

The collocation yields 6\,629 independent L2B matches from the four
satellite instruments (Table~\ref{tab:datasets}), distributed across
26~buoys. Scatterometer winds are quality-screened below
$\sim$3~m~s$^{-1}$, so the matched winds span 3--21~m~s$^{-1}$. ASCAT
instruments contribute 2\,699 collocations on the 12.5-km grid, while
the HY-2 instruments contribute 3\,930 on the 25-km grid. The mean
distance from a MELODI session to the nearest matched scatterometer cell
is 8.5~km, and the mean temporal offset is 15.1~minutes. This representativeness mismatch---a point
measurement compared against a satellite cell of 12.5--25-km grid spacing---sets
an irreducible error floor of approximately 0.7--1.0~m~s$^{-1}$ for wind
speed comparisons \citep{Stoffelen1998}, which must be considered when
interpreting the validation statistics in section~\ref{sec:results}.

\subsection{Sofar Spotter reference}\label{sec:spotter_data}

The Sofar Spotter is the established commercial drifter for wave-based wind
retrieval and serves as an independent reference for the MELODI product. The
Spotter~v3 is a compact (42~cm diameter), solar-powered buoy that measures
surface displacement directly from multi-constellation GNSS tracking at
2.5~Hz; unlike the accelerometer-based MELODI it avoids the double-integration
step and its associated low-frequency noise. Each nominally hourly cycle yields
a directional wave spectrum (39~frequency bins over 0.029--0.654~Hz, 120~azimuthal
bins) and operationally derives wind speed and direction from the
equilibrium range of that spectrum using the single-band inversion of
\citet{Voermans2020}, now deployed across a global fleet \citep{Dorsay2023}.
As the most widely used drifter wind solution, it is the natural benchmark for
our retrieval. During the OTC25 campaign one Spotter (SPOT-32265C) drifted
alongside the recovered MELODI buoy OTC25\_16 for 14~days (14--28~May~2025)
in the northeast Atlantic ($\sim$55$^\circ$N, 17$^\circ$W), within $<$30~km for
the first week, providing 665 collocated hourly samples; this co-deployment
supports the direct intercomparison in section~\ref{sec:spotter}.

\section{Data processing}\label{sec:processing}

\begin{figure*}[t]
\centering
\includegraphics[width=\textwidth]{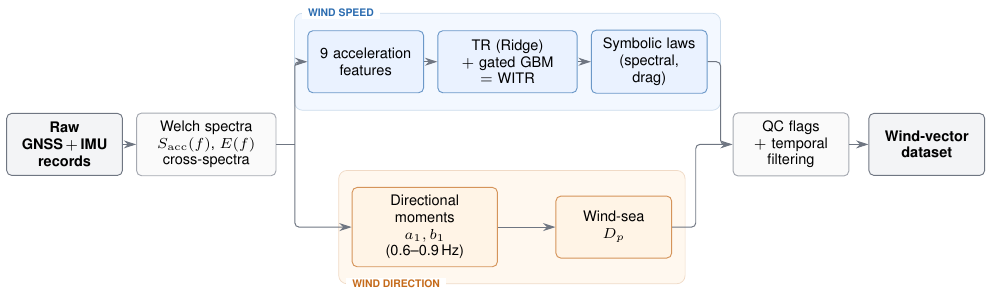}
\caption{Processing chain from raw drifter records to the assembled
  wind-vector dataset. The wind-speed branch (blue) extracts
  acceleration-spectrum features, maps them to speed with the
  Tikhonov-regularized model plus gated GBM correction (WITR), and distills
  the result into closed-form analytical laws; the wind-direction branch
  (orange) forms the wind-sea directional moments and peak direction $D_p$.
  Both are quality-flagged and temporally filtered before assembly.}
\label{fig:pipeline}
\end{figure*}

The processing rests on one physical principle: the equilibrium range of
the wind-wave spectrum encodes wind speed, while the directional
distribution of the high-frequency waves carries wind direction.
Figure~\ref{fig:pipeline} summarizes the chain from raw records to the
assembled wind vector.

\subsection{Omnidirectional spectrum}

Omnidirectional displacement spectra are estimated by a modified Welch
method \citep{Welch1967}: each 22-min session is split into 256-s segments
(75\% overlap, $\sim$17 effective), sine-tapered with variance correction,
periodogram-averaged, and merged in three-bin groups to a resolution of $\sim$0.012~Hz.

For Tier~A buoys, the onboard-computed acceleration spectrum
$S_{\mathrm{acc}}(f)$ is converted to displacement spectral density via
Eq.~(\ref{eq:acc2disp}). For Tier~B buoys with recovered AHRS data, the
vertical acceleration is first despiked (10-$\sigma$ threshold) and
high-pass filtered using a first-order infinite-impulse-response (IIR) filter with
$\tau_{\mathrm{RC}} = 3.5$~s (cutoff $\sim$0.045~Hz) before spectral
estimation. For Tier~C buoys, displacement spectra are computed from GNSS
horizontal velocities via $S_\eta(f) = S_v(f)/(2\pi f)^2$. In all
cases, a half-cosine low-frequency taper is applied between 0.025 and
0.04~Hz to suppress noise amplification from the $f^{-4}$ or
$f^{-2}$ transfer functions at long periods; the taper carries no
wave information---it exists so that the integral moments and band levels
computed downstream are not corrupted by amplified low-frequency noise.

The tapered spectra retain useful energy over 0.035--1.0~Hz, the analysis
band for all feature extraction; the four-band Toba parameters and
single-band inversion of section~\ref{sec:theory} are evaluated on it, and
the wind-sensitive features are extracted from it (section~\ref{sec:feat});
Fig.~S1 in the online supplemental material illustrates this processing on
two contrasting sessions from the OTC25\_16 buoy.

\subsection{Directional spectrum from IMU}\label{sec:directional}

For Tier~B and C buoys, directional wave information is extracted from the
IMU gyroscope and magnetometer channels. The body-frame roll and pitch
\emph{angles} $\theta_x,\theta_y$ are first recovered by integrating the
gyroscope $x$ and $y$ angular rates (then high-pass filtered, fourth-order
Butterworth, 0.02~Hz cutoff, to remove integration drift). Under the
small-slope approximation the sea-surface slope equals the buoy tilt
angle, $\partial\eta/\partial x \approx \tan\theta \approx \theta$ (rad,
dimensionless), so the geographic east and north slope components follow
by rotating the body-frame angles through the buoy heading $\psi$:
\begin{equation}\label{eq:slopes}
\frac{\partial\eta}{\partial x} = \theta_y\cos\psi
  + \theta_x\sin\psi \,,\quad
\frac{\partial\eta}{\partial y} = \theta_y\sin\psi
  - \theta_x\cos\psi \,,
\end{equation}
where $\theta_x,\theta_y$ are the integrated, high-pass-filtered roll and
pitch angles and $\psi$ is the buoy heading. This is the standard
pitch--roll slope formulation of heave--tilt directional-wave analysis. The heading is
obtained from the AHRS firmware output (Tier~B) or from a complementary
filter fusing magnetometer heading with integrated gyroscope yaw rate
(time constant $\tau = 2.0$~s; Tier~C).

The directional distribution at each frequency is summarized by its first
four Fourier coefficients $a_1,b_1,a_2,b_2$, computed from the auto- and
cross-spectra of the two slope components and the vertical acceleration
\citep{Kuik1988}:
\begin{equation}\label{eq:a1b1}
\begin{aligned}
a_1(f) &= \frac{\mathrm{Im}[C_{s_x z}(f)]}
  {\sqrt{[P_{s_x}(f) + P_{s_y}(f)]\,P_z(f)}} \,, \\[4pt]
b_1(f) &= \frac{\mathrm{Im}[C_{s_y z}(f)]}
  {\sqrt{[P_{s_x}(f) + P_{s_y}(f)]\,P_z(f)}} \,, \\[4pt]
a_2(f) &= \frac{P_{s_x}(f) - P_{s_y}(f)}{P_{s_x}(f) + P_{s_y}(f)} \,, \\[4pt]
b_2(f) &= \frac{2\,\mathrm{Re}[C_{s_x s_y}(f)]}{P_{s_x}(f) + P_{s_y}(f)} \,,
\end{aligned}
\end{equation}
where $P_{s_x}$, $P_{s_y}$, $P_z$ are the auto-spectra of the east slope,
north slope, and vertical acceleration, and $C_{s_x z}$, $C_{s_y z}$,
$C_{s_x s_y}$ the corresponding cross-spectra. The first-order pair
$(a_1,b_1)$ fixes the mean wave direction
$\theta_p(f) = \arctan\!2(b_1, a_1)$ (mapped to the compass convention)
and the directional coherence $r_1(f) = \sqrt{a_1^2 + b_1^2}$, while the
second-order pair $(a_2,b_2)$ constrains the angular width of the
directional lobe.

The retrieval uses these directional moments directly; the full
two-dimensional spectrum $S(f,\theta)$ is not reconstructed. The moments
are filtered before use---a noise floor at 1\% of peak energy, 5-band
running-mean smoothing, and clipping to the physical range $[-1, 1]$.

The directional coherence $r_1$ in the high-frequency wind-sea band
(0.6--0.9~Hz) serves as a quality metric: values of $r_1 \geq 0.4$
indicate a well-defined wind-sea direction, while $r_1 < 0.2$ flags
sessions with ambiguous directionality (e.g., during swell--wind-sea
coupling transitions). For Tier~B data (AHRS at 3.2~Hz), the typical
dominant first-order moment in the wind-sea band is $|a_1| \approx 0.65$; Tier~C data
(raw IMU at 100~Hz) achieves $|a_1| \approx 0.85$
(Fig.~\ref{fig:dir_coeffs}).

\subsection{High-frequency response}

MELODI resolves waves up to 1.0~Hz, beyond the range of most operational
wave buoys (typically 0.5--0.6~Hz); as discussed in
section~\ref{sec:theory}, these locally generated high-frequency waves
equilibrate with the wind within minutes and are therefore the most
informative bands for instantaneous wind retrieval.

The retrieved wind-speed series is post-processed with a two-stage temporal filter
applied independently to each session sequence: a spike test removes
isolated outliers (threshold four times the median absolute deviation),
followed by a short Savitzky--Golay smoother
(window~5, order~2). This smoother, like the directional one of section~\ref{sec:dir_filter},
uses a centered window and is therefore applied in post-processing.

\subsection{Reference collocation}\label{sec:collocation}

The reference wind products are collocated with the MELODI sessions as
follows.

\textit{ERA5 reanalysis.} ERA5 10-m winds are bilinearly interpolated from
the native 0.25$^{\circ}$ fields to each buoy position and linearly in time
to each session (ERA5T near-real-time fields for 2025). ERA5 serves as a
secondary reference; the scatterometer winds are the primary one.

\textit{Scatterometers.} The scatterometer is the higher-resolution
reference (the temporal criteria are the same $\pm$30~min for both
references; the substantive difference is spatial resolution). For each
overpass the swath cells are screened to
reject non-finite and rain-contaminated retrievals (rain being especially
important for the Ku-band HY-2 instruments). Cells within a great-circle
distance $d\leq25$~km and $|\Delta t|\leq30$~min of a session are retained;
the matched wind speed is their arithmetic mean and the direction their
circular mean. All directions are placed on the common meteorological
convention (direction \emph{from} which the wind blows) before comparison.
When two sessions match the same overpass, the closer one is kept.

Table~\ref{tab:subsets} reconciles the buoy tiers and collocation subsets.
The operational product is the 28 buoys returning retrievable data (24
Tier-A, speed only; 4 Tier-B, speed and direction); the four Tier-C buoys
(N1--N4) are a separate collocated test deployment, not part of the product.
The full L2B set (6\,629 collocations, 26 buoys) trains the spectrum-only
WITR; the tight-QC subset ($\leq$8~km, $\leq$25~min; 2\,433, 25 buoys)
drives the leave-one-buoy-out cross-validation, with per-buoy errors
reported for the 17 buoys carrying $\geq$10 collocations; and the
motion-based reduced drag law is restricted to the four recovered buoys
(4\,437 / 1\,942 collocations).

\begin{table*}[t]
\caption{Dataset and collocation hierarchy: buoy counts, sample sizes, and
  the retrieval evaluated on each. The operational product is Tier~A~+~B
  (28 buoys); Tier~C (N1--N4) is a separate collocated test deployment, not
  part of the product. ``Recovered'' denotes the four Tier-B buoys that
  carry IMU motion data.}
\label{tab:subsets}
\begin{center}
\begin{tabular}{lrrl}
\hline\hline
Subset & Buoys & $N$ & Used by \\
\hline
Operational product (Tier~A\,+\,B)    & 28 & 29\,775 & speed (all); direction (Tier~B) \\
\quad Tier~A (satellite spectra)      & 24 & 8\,033  & wind speed only \\
\quad Tier~B (recovered IMU)          & 4  & 21\,742 & speed + direction; feature dev. \\
Tier~C (GNSS+IMU test)                & 4  & 101     & directional cross-check (not in product) \\
\hline
ERA5 collocations (Tier~B)            & 4  & 21\,742 & feature dev.; ERA5 validation \\
L2B scatterometer collocations        & 26 & 6\,629  & WITR (full L2B) \\
\quad recovered-buoy subset           & 4  & 4\,437  & reduced drag law (L2B) \\
Tight-QC scat ($\leq$8~km, $\leq$25~min) & 25 & 2\,433 & WITR LOBO; spectral laws \\
\quad per-buoy LOBO reported          & 17 & ---     & per-buoy speed errors \\
\quad recovered-buoy subset           & 4  & 1\,942  & reduced drag law (tight-QC) \\
Direction collocations (ASCAT L2)     & 4  & 521     & wind-direction validation \\
\hline\hline
\end{tabular}
\end{center}
\end{table*}

\section{Wind speed retrieval}\label{sec:speed}

Wind speed is retrieved in a sequence of three stages of increasing
physical transparency, built on a common feature basis. We first extract
a compact set of acceleration-spectrum features
(section~\ref{sec:feat}). A Tikhonov-regularized linear model
(\emph{TR}; section~\ref{sec:tr}) then maps these features to wind speed,
providing the baseline inversion. The linear model carries a
systematic residual bias---most notably the saturation of
equilibrium-range energy at strong winds---which is removed by a
sigmoid-gated nonlinear correction. The corrected model
(section~\ref{sec:tr_rect}) is referred to as WITR (Wind Inversion using
Tikhonov Regularization). Both stages are trained against the
scatterometer winds, with ERA5 reanalysis used during feature
development.

\subsection{Feature extraction and selection}\label{sec:feat}

From each MELODI session, a broad set of candidate spectral features is
extracted to characterize the wind-sea state. The initial candidate pool
comprises approximately 52 features organized into five categories:
band-level spectral features ($\beta_4$, spectral slope, level,
and Toba wind estimate in each of the four bands; 16 features),
acceleration spectrum features (five fine-band mean levels between 0.12 and
0.70~Hz, slopes, moments, and frequency percentiles; $\sim$20 features),
global spectral shape ($H_s$, $f_p$, spectral width; 8 features),
high-frequency noise floor and signal-to-noise ratios (5 features), and
environmental variables (SST, drift speed, $\Delta T$; 3 features).
The full candidate list, with each feature's mathematical definition and
its ranked importance in a preliminary Tikhonov-regularized model, is
provided with the archived model code (data-availability statement).

Backward elimination using LOBO cross-validation
against scatterometer wind speeds reduces this pool to a minimal set of
9 acceleration-spectrum features (the relative performance of the reduced and
full feature sets is quantified in section~\ref{sec:results}). The
selected WITR features are:
\begin{enumerate}
\item Five acceleration band-mean levels: $\overline{S}_{\mathrm{acc}}$, the
  mean of $S_{\mathrm{acc}}(f)$ over each band, in the bands 0.12--0.18,
  0.18--0.25, 0.25--0.35, 0.35--0.50, and 0.50--0.70~Hz.
\item The high-frequency noise floor at 0.60--0.80~Hz.
\item Acceleration spectral slopes in 0.25--0.50 and 0.50--1.00~Hz.
\item The 25th-percentile cumulative frequency $f_{25}$.
\end{enumerate}
These features are all derived from the acceleration spectrum
$S_{\mathrm{acc}}(f)$ and require no conversion to displacement, no
motion sensor data (pitch, roll), and no environmental inputs (SST).
This makes the WITR model applicable uniformly to all three data tiers,
including satellite-transmitted Tier~A spectra.

We note that the feature set itself was selected on the full collocation
corpus, so the reported RMSD may be mildly optimistic despite the
buoy-level holdout; a fully nested cross-validation with inner-loop
feature selection is left for future work.

\begin{table}[t]
\caption{WITR model parameters.}\label{tab:features}
\centering
\footnotesize
\setlength{\tabcolsep}{3pt}
\begin{tabular}{@{}ll p{0.92in}@{}}
\hline\hline
Parameter & Value & Description \\
\hline
\multicolumn{3}{@{}l}{\textit{Feature selection}} \\
Candidate features & 52 & spectral + environ.\ pool \\
Final WITR features & 9 & acc.\ spectrum only \\
Selection method & backward elim. & LOBO vs scat. \\
Cand.\ motion feats. & $\sim$11 & pitch, roll, $\sigma_{a_z}$, yaw (not in WITR) \\
Reduced-drag inputs & 2 & $\sigma_{a_z}$, pitch (recovered only) \\
\hline
\multicolumn{3}{@{}l}{\textit{Tikhonov regularization (TR)}} \\
Regularization $\alpha$ & 1.0 & L2 penalty \\
Feature scaling & StandardScaler & zero mean, unit var. \\
Cross-validation & LOBO & leave-one-buoy-out \\
\hline
\multicolumn{3}{@{}l}{\textit{Regime gating (sigmoid)}} \\
High-wind threshold & 9.0 & sigmoid centre (m~s$^{-1}$) \\
Transition width & 1.5 & sigmoid width (m~s$^{-1}$) \\
High-wind upweighting & 2--5$\times$ & $\geq$10 to $\geq$14~m~s$^{-1}$ \\
\hline
\multicolumn{3}{@{}l}{\textit{Analytical fitting (symbolic regression)}} \\
Max complexity & 50 / 12 & nodes (full/simple) \\
Iterations & 2000 & evol.\ cycles \\
Operators & $+,-,\times,\div$ & binary \\
 & $\sqrt{\cdot},\log,|\cdot|,(\cdot)^2$ & unary \\
Parsimony penalty & 0.003 / 0.01 & complexity cost \\
Populations & 40$\times$50 & parallel search \\
\hline\hline
\end{tabular}
\end{table}

\subsection{Tikhonov regression (TR)}\label{sec:tr}

The baseline retrieval is a Tikhonov-regularized (Ridge) linear model
\citep{Hoerl1970} mapping the 9 selected acceleration features
$\mathbf{X}$ to scatterometer wind speed $\mathbf{y}$, with coefficients
minimizing
\begin{equation}\label{eq:tr}
\boldsymbol{\hat{\beta}} = \arg\min_{\boldsymbol{\beta}}
\left\{ \|\mathbf{y} - \mathbf{X}\boldsymbol{\beta}\|_2^2
+ \alpha\,\|\boldsymbol{\beta}\|_2^2 \right\} \,,
\end{equation}
where $\alpha = 1.0$ is the regularization parameter, selected by
cross-validation. Features enter the model as their raw physical values and are
standardized to zero mean and unit variance before fitting. Table~\ref{tab:v14_coeffs} lists the fitted coefficients and the
standardization constants needed to apply them. The dominant predictors
are the mid-band acceleration levels and the 25th-percentile frequency,
and the table allows the linear stage to be evaluated by hand.

\input{tables/v14_coeffs_v3}

\subsection{Nonlinear correction stage}\label{sec:tr_rect}

The linear TR stage cannot represent the nonlinear dependencies of the
wind--wave relationship, particularly the saturation of equilibrium-range
energy at high winds, and consequently underestimates strong winds. This
residual bias is corrected by a second, nonlinear stage. A
gradient-boosting machine (GBM; a decision-tree ensemble, 150~iterations, max depth 3, learning
rate 0.05, L2 regularization 2.0) is trained on the TR residuals
$\mathbf{y} - \hat{\mathbf{y}}_{\mathrm{TR}}$ against the same
scatterometer reference.
Strong-wind samples ($\geq$10~m~s$^{-1}$) are upweighted during GBM training
(2$\times$ at $\geq$10, 3$\times$ at $\geq$12, 5$\times$ at
$\geq$14~m~s$^{-1}$). The final prediction blends the two stages via a
sigmoid gate:
\begin{equation}\label{eq:sigmoid}
\begin{split}
\hat{U}_{10} &= \hat{U}_{\mathrm{TR}} + w_s\,\hat{U}_{\mathrm{GBM}} \,, \\
w_s &= \frac{1}{1+\exp\!\left[-(\hat{U}_{\mathrm{TR}}-c)/\sigma\right]} \,,
\end{split}
\end{equation}
where $c = 9.0$~m~s$^{-1}$ and $\sigma = 1.5$~m~s$^{-1}$ are the
sigmoid center and width. The GBM correction is progressively activated
as the linear prediction exceeds $c$, providing a smooth transition
from the purely linear regime at moderate winds to the nonlinear
correction at strong winds. The final prediction is clipped to
$[0,\,35]$~m~s$^{-1}$.

The gated prediction is completed by an affine variance-matching
calibration, $\hat{U}_{10}^{\mathrm{cal}} = \bar{U}_{\mathrm{scat}} +
s\,(\hat{U}_{10} - \bar{\hat{U}}_{10})$ with $s =
\sigma_{U_{\mathrm{scat}}}/\sigma_{\hat{U}_{10}} = 1.054$ fitted on the
training collocations. Least-squares regression on noisy features
attenuates the retrieved dynamic range (the fitted response slope is
biased below unity), over-predicting the lightest and under-predicting
the strongest winds \citep{Stoffelen1998}; the calibration restores a
unit response slope at negligible cost in overall RMSD.

The model is trained directly against scatterometer winds
(section~\ref{sec:scat}), not ERA5, and evaluated by leave-one-buoy-out
cross-validation: each buoy is predicted by a model trained on the
others, testing geographic generalization across basins. Because the
scatterometer is also the training target, it is this buoy-level holdout
that guards the reported accuracy against in-sample optimism.

\subsection{Analytical distillation}\label{sec:analytical}

WITR serves here as a supervised \emph{teacher} model: it learns the mapping
between acceleration-spectrum shape and scatterometer wind speed and marks
the best accuracy we obtain from the available feature set. We now distill
this trained model into closed-form equations with symbolic regression
\citep[PySR;][]{Cranmer2023}. The resulting expressions are transparent,
dimensionally interpretable, and cheap enough to run onboard; the
algorithm searches for equations that balance prediction accuracy against
expression complexity (number of nodes in the expression tree).

The distillation target is the WITR model prediction (not the ERA5
reference directly), so that the symbolic search discovers the structure
learned by the regularized inversion without being confounded by
representativeness noise. The input feature set comprises $\sim$35
variables: the physics-first spectral
features, top predictors of the WITR model selected by Shapley additive explanations (SHAP), and eight
composite features encoding known physical relationships (Toba proxy
$\sqrt{H_s f_p}$, stress proxy $H_s f_p^2$, roughness length
$z_0 = u_*^2/g$, and the explicit Toba inversion $U^{\mathrm{Toba}}$
with iterative drag). The constants of the closed-form laws below
incorporate the same final variance-matching normalization as WITR
(section~\ref{sec:tr_rect}).

The search yields a hierarchy of equations of increasing complexity, of
which we retain four:

\textit{Proportional scaling (complexity 3).} The simplest useful
equation is a direct proportionality between wind speed and the zeroth
moment of the acceleration spectrum:
\begin{equation}\label{eq:pysr_c3}
U_{10} = 7.1\,m_0^{\mathrm{acc}} \,.
\end{equation}
This captures the bulk relationship but lacks regime sensitivity.

\textit{Spectral law (the principal spectrum-only equation).} A wave-age
correction to the standard Toba inversion, requiring only the
omnidirectional acceleration spectrum, yields a substantial improvement:
\begin{equation}\label{eq:pysr_c8}
U_{10} = U^{\mathrm{Toba}}_{\mathrm{MID}}
\!\left(0.257 + 0.0178\,U^{\mathrm{Toba}}_{\mathrm{LO}}\right)
+ 2.13 \,,
\end{equation}
where $U^{\mathrm{Toba}}_{\mathrm{MID}}$ and
$U^{\mathrm{Toba}}_{\mathrm{LO}}$ are Toba inversions from the MID and
LO bands [Eq.~(\ref{eq:toba_inversion})]. The LO term acts as a wave-age
correction: young, actively growing seas have elevated LO-band energy
relative to equilibrium, amplifying the MID-band estimate. This
two-variable form requires only the omnidirectional acceleration spectrum
yet captures the leading wave-age dependence in analytically transparent
form, attaining RMSD $1.09$~m~s$^{-1}$ against the scatterometer winds
($1.19$ against ERA5).

\textit{Extended spectral law.} Adding the high-frequency
band through a spectral-equilibrium-deficit term yields
\begin{equation}\label{eq:pysr_c15}
\begin{split}
U_{10} = {}& 0.418\,U^{\mathrm{Toba}}_{\mathrm{MID}} + 1.31 \\
& + 0.00935\!\left({U^{\mathrm{Toba}}_{\mathrm{LO}}}^2
+ \big(U^{\mathrm{Toba}}_{\mathrm{LO}} - U^{\mathrm{Toba}}_{\mathrm{HI}}\big)^2\right) ,
\end{split}
\end{equation}
where the squared LO$-$HI difference measures the departure from spectral
equilibrium (active wind-sea growth or swell decay). This improves to RMSD
$1.02$~m~s$^{-1}$ against the scatterometer winds ($1.16$ against ERA5)
while still using only the omnidirectional spectrum, so it remains
applicable to all data tiers including the satellite-transmitted Tier~A.

\textit{Reduced drag law (with buoy motion; the principal analytical
retrieval).} This law draws additionally on the internal buoy-motion measurements
($\sigma_{a_z}$ and pitch RMS) beyond the spectrum alone. Run with
PySR's dimensional-consistency constraint over Buckingham-$\pi$ groups,
the symbolic search discovers a compact dimensionless drag closure:
\begin{equation}\label{eq:reduced_drag}
\begin{split}
\frac{U_{10}}{u_*} = {}& 26.9 - \frac{0.829}{\sigma_{a_z}/g}
+ \left\lvert \frac{68.3}{\sqrt{\mathrm{Re}_* + 0.46}} - 3.11\,r \right\rvert \\
& - \frac{0.310}{\theta} - 0.00243\,\mathrm{Re}_* \,,
\end{split}
\end{equation}
where $u_*$ is the friction velocity from $\beta_4$(MID) via
Eq.~(\ref{eq:toba_inversion}), $\sigma_{a_z}/g$ is the vertical-acceleration
standard deviation normalized by gravity $g$,
$\mathrm{Re}_* = u_*^3/(g\,\nu_a)$ is the
dimensionless roughness Reynolds number
($\nu_a = 1.5 \times 10^{-5}$~m$^2$~s$^{-1}$), $r =
u_{*,\mathrm{LO}}/u_{*,\mathrm{HI}}$ is the spectral-maturity ratio, and
$\theta$ is the buoy pitch RMS in radians. All four groups are
dimensionless, so the law is dimensionally consistent by construction.

We call it the \emph{reduced} drag law for a second, physical reason: the
leading constant 26.9 is the canonical neutral log-law value of
$U_{10}/u_*$ ($\approx$25--30; \citealp{Smith1980}), and the high-wind
term $-0.00243\,\mathrm{Re}_*$ (which grows as $u_*^3$) bends the implied
drag coefficient over so that it peaks near 8--11~m~s$^{-1}$
($C_d \approx 3.3\times10^{-3}$) and then \emph{decreases}
($C_d \approx 2.3\times10^{-3}$ by 14--18~m~s$^{-1}$). This turnover
occurs well below the winds at which the physical drag coefficient is
observed to saturate \citep{Powell2003,Donelan2004}, so we read the term
as an empirical strong-wind correction fitted on a sparse tail rather
than as recovery of the physical drag rollover. The sea-state term
$-0.829/(\sigma_{a_z}/g)$ and the viscous--maturity term
$|68.3/\sqrt{\mathrm{Re}_*+0.46} - 3.11\,r|$ encode, respectively, the
energy level and the smooth-to-rough/maturity transition. The reduced drag
law is most valuable at the extremes: against the scatterometer winds it
attains a leave-one-buoy-out RMSD of 0.94~m~s$^{-1}$ (0.93 in sample) and
cuts the strong-wind ($>$12~m~s$^{-1}$) RMSD to 1.28~m~s$^{-1}$, at the cost
of requiring the IMU motion inputs available only for the recovered buoys.

The convergence of these structures across independent runs with
different initializations indicates a reproducible empirical structure,
although its physical interpretation remains provisional. The reduced drag law
[Eq.~(\ref{eq:reduced_drag})] uses $\beta_4$ in three bands (giving $u_*$
and the maturity ratio $r$) together with $\sigma_{a_z}$ and pitch
RMS---four dimensionless groups---requires $\sim$20~floating-point
operations, and is fully implementable on the buoy's onboard
microcontroller.

The two inverse terms in Eq.~(\ref{eq:reduced_drag}),
$1/(\sigma_{a_z}/g)$ and $1/\theta$, become ill-conditioned at very low
wave energy or very small pitch RMS. In operational use Eq.~(\ref{eq:reduced_drag})
is therefore evaluated only when $\sigma_{a_z}/g$ and $\theta$ exceed the
empirical floors used during training; otherwise the spectrum-only law
[Eq.~(\ref{eq:pysr_c15})] or WITR is used and a low-confidence flag is
assigned.

\section{Wind direction retrieval}\label{sec:direction}

Unlike wind speed, which is derived statistically from spectral energy
levels, wind direction is estimated from the directional properties of
the high-frequency wind-sea using a physics-based approach that requires
no training data.

\subsection{Windsea directional moments}\label{sec:dir_moments}

The wind direction is inferred from the peak direction of
high-frequency wind waves, which are generated locally by the wind and
align with the wind vector within minutes. The key point is simple: a band chosen well above the swell peak
isolates the locally generated wind sea, whose propagation direction
follows the current wind.

The directional Fourier coefficients $a_1(f)$ and $b_1(f)$ are computed
from IMU-derived surface slopes and vertical acceleration as described
in section~\ref{sec:directional} [Eq.~(\ref{eq:a1b1})]. For wind
direction retrieval, these moments are averaged over a wind-sea
frequency band with energy weighting:
\begin{equation}\label{eq:dir_avg}
\bar{a}_1 = \frac{\sum_{f \in \mathrm{band}} a_1(f)\,P_z(f)}
  {\sum_{f \in \mathrm{band}} P_z(f)} \,,\quad
\bar{b}_1 = \frac{\sum_{f \in \mathrm{band}} b_1(f)\,P_z(f)}
  {\sum_{f \in \mathrm{band}} P_z(f)} \,,
\end{equation}
and the wind-sea peak direction is
\begin{equation}\label{eq:dp_windsea}
D_p = \left(90^{\circ} - \arctan\!2(\bar{b}_1,\,\bar{a}_1)\right)
\bmod{360^{\circ}} \,,
\end{equation}
in compass convention (clockwise from north, direction from which the
wind blows).

\begin{figure*}[t]
  \centerline{\includegraphics[width=\textwidth]{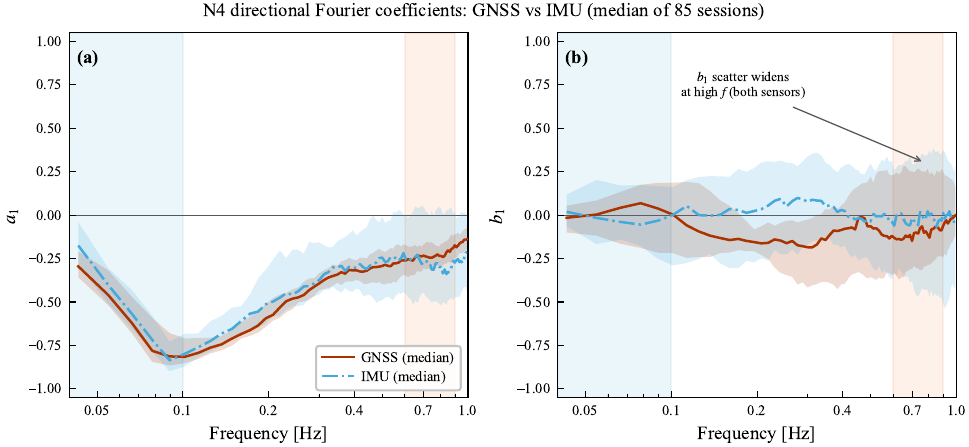}}
  \caption{First-order directional Fourier coefficients $a_1,b_1$ versus frequency (median and 10--90\% envelope over the 85 collocated N4 Tier-C sessions), comparing GNSS and IMU in fixed geographic east--north axes; the second-order pair is shown in Fig.~S2 of the online supplemental material. The GNSS and IMU medians track each other across the band; the envelopes widen at high frequency for both sensors. During this deployment the mean wave direction lay nearly along the cosine axis, so $a_1$ carries the directional signal while $b_1$ scatters about zero. The wind-direction retrieval uses only the first-moment direction fixed by $(a_1,b_1)$; the IMU alone suffices for wind-sea direction.}
  \label{fig:dir_coeffs}
\end{figure*}

The mapping from the truncated Fourier moments to the directional
distribution is standard \citep{LonguetHiggins1975,Kuik1988}: the full
spectrum $D(f,\theta)$ uses all four coefficients $(a_1,b_1,a_2,b_2)$, but
the \emph{mean} wave direction and its coherence are fixed by the
first-order pair $(a_1,b_1)$ alone, the second pair $(a_2,b_2)$ only setting
the angular width of the lobe \citep{Kuik1988,LygreKrogstad1986}. The
second-order pair therefore does not enter the retrieval, and the
precision of $(a_2,b_2)$ is immaterial for wind direction (they are shown in Fig.~S2 of the online
supplemental material). Figure~\ref{fig:dir_coeffs} compares the
coefficients from the GNSS-velocity and IMU heave-tilt methods on the
collocated N4 (Tier-C) deployment; the axes are fixed geographic
east--north, with no record-dependent rotation. The GNSS and IMU medians
track each other across the band, and the inter-session envelopes widen
at high frequency for both sensors. During this deployment the mean
wave direction lay nearly along the cosine axis, so $a_1$ carries the
directional signal while $b_1$ scatters about zero---a property of the
geometry, not of the sensors. The derived first-moment direction remains
stable after coherence screening and temporal filtering, as confirmed by
the scatterometer validation (section~\ref{sec:dir_val}).

On N4, where both sensors are available, the IMU and GNSS $(a_1,b_1)$
directions agree to within $-1.8^{\circ}$ in the mean (standard deviation
$8^{\circ}$) across the energetic swell band, so the IMU heave-tilt
estimate recovers the same first-moment direction as the GNSS orbital
velocity---the configuration used by the recovered Tier~B buoys, which
lack high-rate GNSS wave records. Because N4 itself drifted under swell with little wind-sea energy,
the decisive wind-sea-band test is the collocation of the IMU-only $D_p$
against scatterometer winds (Fig.~\ref{fig:dir_scatter}; MAAD
$8.3^{\circ}$).

The choice of frequency band is critical. The standard approach of using
the 0.35--0.60~Hz band suffers from swell contamination in mixed sea
states, where long-period swell energy leaks into the directional
moments and biases the retrieved direction. We instead use the
0.60--0.90~Hz band, where the higher frequencies are dominated by locally
generated wind-sea and are much less susceptible to swell contamination; the
resulting reduction in direction difference relative to the conventional
0.35--0.60~Hz choice is quantified in section~\ref{sec:results}
(Table~\ref{tab:band_sens}). As with the speed model, the band limits, coherence threshold, and
filtering parameters were tuned on the same scatterometer collocations
used to report performance. The direction retrieval contains no fitted
regression, so this tuning amounts to a handful of discrete choices;
even so, the reported direction differences should be considered mildly
optimistic rather than fully out-of-sample.

The directional coherence $r_1 = \sqrt{\bar{a}_1^2 + \bar{b}_1^2}$
quantifies the quality of the directional estimate. Values of
$r_1 \geq 0.4$ indicate a well-defined unimodal wind-sea, while
$r_1 < 0.2$ signals ambiguous conditions such as swell--wind-sea
coupling transitions or rapidly turning winds.

\subsection{Filtering and quality control}\label{sec:dir_filter}

The raw session-by-session $D_p$ series carries isolated outliers from
transient coherence drops, heading-calibration errors, or brief swell
intrusions. These are removed by a multi-stage pipeline operating on
$(\sin D_p, \cos D_p)$ to respect the $0^{\circ}/360^{\circ}$ wrap:
iterative Savitzky--Golay outlier detection (window~11, order~2,
$35^{\circ}$ threshold, three passes) and consecutive-jump flagging,
linear interpolation of the flagged points, a final Savitzky--Golay smooth
(window~5), and a two-pass median despike. Applied twice per buoy, it
yields a continuous, 100\%-coverage series while preserving genuine
direction changes.

The filter's effect is confined to sub-synoptic lags: the lag-1
($\approx$30-min) circular autocorrelation of the $D_p$ series rises from
0.96 (raw) to 0.99 (filtered) and the two converge by $\approx$8~h (0.71
versus 0.73), so the smoothing suppresses session-scale jitter
(session-to-session spread 17$^{\circ}$ to 6$^{\circ}$) without borrowing
information across the synoptic scales sampled by the validation. At the
scatterometer overpasses the raw, unfiltered $D_p$ already attains a
MAAD of 11.1$^{\circ}$; the filtering improves this to
8.8$^{\circ}$.

Sessions with $r_1 < 0.2$ in the wind-sea band are flagged as reduced
confidence. These
sessions are retained in the product but may be filtered by downstream
applications. Low-coherence conditions are rare: pooled over the
21\,742 recovered-buoy sessions, the wind-sea $r_1$ has median 0.43
(10th--90th percentiles 0.32--0.56); 63\% of sessions exceed the
well-defined threshold of 0.4 and only 1.7\% fall below 0.2
(0.4--5.0\% per buoy; Table~\ref{tab:band_sens}; the full distribution
is shown in Fig.~S3 of the online supplemental material).

\section{Results and validation}\label{sec:results}

The retrievals are now evaluated against held-out scatterometer
collocations and against ERA5. The direction retrieval uses no fitted
reference-wind regression at inference time; its band and filtering
choices were, however, selected using the scatterometer collocations, so
the reported directional differences are mildly optimistic rather than
fully out of sample. Throughout, statistics of the difference between two
collocated systems are reported as root-mean-square differences (RMSD),
since both systems carry measurement and representativeness errors;
estimates of the error of a single system relative to the unknown truth,
as obtained from triple collocation (section~\ref{sec:accfloor}), are
the only quantities reported as errors. Wind speed comes first, through
the three principal retrievals in order of increasing physical
transparency (the classic Toba baseline, WITR, and the reduced drag law),
followed by wind direction, the combined wind-vector time series, and an
error summary.

\subsection{Speed validation}\label{sec:speed_val}

Table~\ref{tab:speed_val} summarizes the wind speed validation against
both ERA5 reanalysis and the scatterometer collocations (by
leave-one-buoy-out cross-validation, since the speed model is trained on
the scatterometer winds) pooled across all five tight-QC scatterometer
products, spanning four satellites (ASCAT MetOp-B/C, HY-2B/C; MetOp-B
contributes both its standard and its coastal L2B products).
For comparison, a traditional single-band Toba inversion
($U^{\mathrm{Toba}}_{\mathrm{LO}}$, on the LO band) is included as a baseline, along
with the four analytical fits of increasing complexity derived in
section~\ref{sec:analytical}.

\begin{table*}[t]
\caption{Wind-speed validation of the retrievals, in order of increasing
  complexity. RMSD in m~s$^{-1}$; $r$ is Pearson correlation; $N_v$ is the
  number of input variables. ``Scat'' is the scatterometer reference
  pooled across all five tight-QC products. The spectral laws use only the
  acceleration spectrum; the reduced drag law additionally requires the
  IMU vertical acceleration and pitch.}\label{tab:speed_val}
\begin{center}
{\small
\begin{tabular}{lcccccc}
\hline\hline
 & & \multicolumn{2}{c}{vs.\ ERA5} & \multicolumn{2}{c}{vs.\ scat (pooled)} & \\
\cline{3-4}\cline{5-6}
Model & $N_v$ & RMSD & $r$ & RMSD & $r$ & Eq.\\
\hline
Toba law (single-band)$^\dagger$ & 1  & 1.82 & 0.88 & 1.80 & 0.88 & (\ref{eq:toba_inversion})\\
Proportional            & 1  & 1.30 & 0.91 & ---  & ---  & (\ref{eq:pysr_c3})\\
Spectral law            & 2  & 1.19 & 0.91 & 1.09 & 0.92 & (\ref{eq:pysr_c8})\\
Extended spectral law   & 3  & 1.16 & 0.92 & 1.02 & 0.93 & (\ref{eq:pysr_c15})\\
\textbf{Reduced drag law} & 4 & 1.17 & 0.92 & \textbf{0.93} & 0.94 & (\ref{eq:reduced_drag})\\
\textbf{WITR}           & 9  & \textbf{1.09} & \textbf{0.93} & \textbf{0.91} & \textbf{0.94} & ---\\
\hline\hline
\end{tabular}}
\end{center}
{\footnotesize
Sample sizes: vs.\ ERA5, $N = 21\,742$ recovered-IMU sessions; vs.\ scat,
$N = 2\,433$ tight-QC collocations (WITR by leave-one-buoy-out, the spectral
laws in closed form), reducing to the $N = 1\,942$ recovered-buoy subset for
the motion-based reduced drag law. The reduced-drag-law scat RMSD is the
closed-form in-sample value (0.93); leave-one-buoy-out gives 0.94, so the
form shows limited buoy-level optimism.
$^\dagger$Toba is the classic single-band equilibrium inversion
[Eq.~(\ref{eq:toba_inversion})] evaluated on the LO band
(0.12--0.30~Hz)---the energetic band closest to the equilibrium range
sampled by published single-band retrievals---with one equilibrium
constant fit on the ERA5 data set (physics-only, no per-buoy
recalibration). The band median of Eq.~(\ref{eq:beta4}) is the level
estimator within any single band, so Eq.~(\ref{eq:toba_inversion})
applies per band.}
\end{table*}

\begin{table}[t]
\caption{Wind-speed RMSD (m~s$^{-1}$) by wind regime against the
  scatterometer reference pooled over all five tight-QC products. Toba, the
  spectral law, and WITR are evaluated on the full 2\,433-collocation
  tight-QC data set; the motion-based reduced drag law on its
  1\,942-collocation recovered-buoy subset. The reduced drag law and WITR
  remain stable at strong winds (RMSD 1.28 and 1.19~m~s$^{-1}$ above
  12~m~s$^{-1}$), whereas the spectrum-only spectral law degrades there.
  The three proposed retrievals agree within $\sim$0.2~m~s$^{-1}$ over
  5--12~m~s$^{-1}$; the single-band Toba inversion is substantially
  noisier throughout and poorest at light winds. Product quality control
  excludes winds below 3~m~s$^{-1}$, so the calm row is effectively
  3--5~m~s$^{-1}$.}
\label{tab:speed_regime}
\centering
\small
\begin{tabular}{lcccc}
\hline\hline
Wind regime & Toba & Spectral & Reduced drag & WITR \\
\hline
0--5~m~s$^{-1}$   & 2.03 & 0.98 & 0.65 & 0.88 \\
5--8~m~s$^{-1}$   & 2.09 & 0.98 & 0.86 & 0.85 \\
8--12~m~s$^{-1}$  & 1.40 & 1.03 & 1.03 & 0.91 \\
$>$12~m~s$^{-1}$  & 1.46 & 1.87 & 1.28 & 1.19 \\
\hline\hline
\end{tabular}
\end{table}

\begin{figure*}[p]
  \centerline{\includegraphics[width=0.88\textwidth]{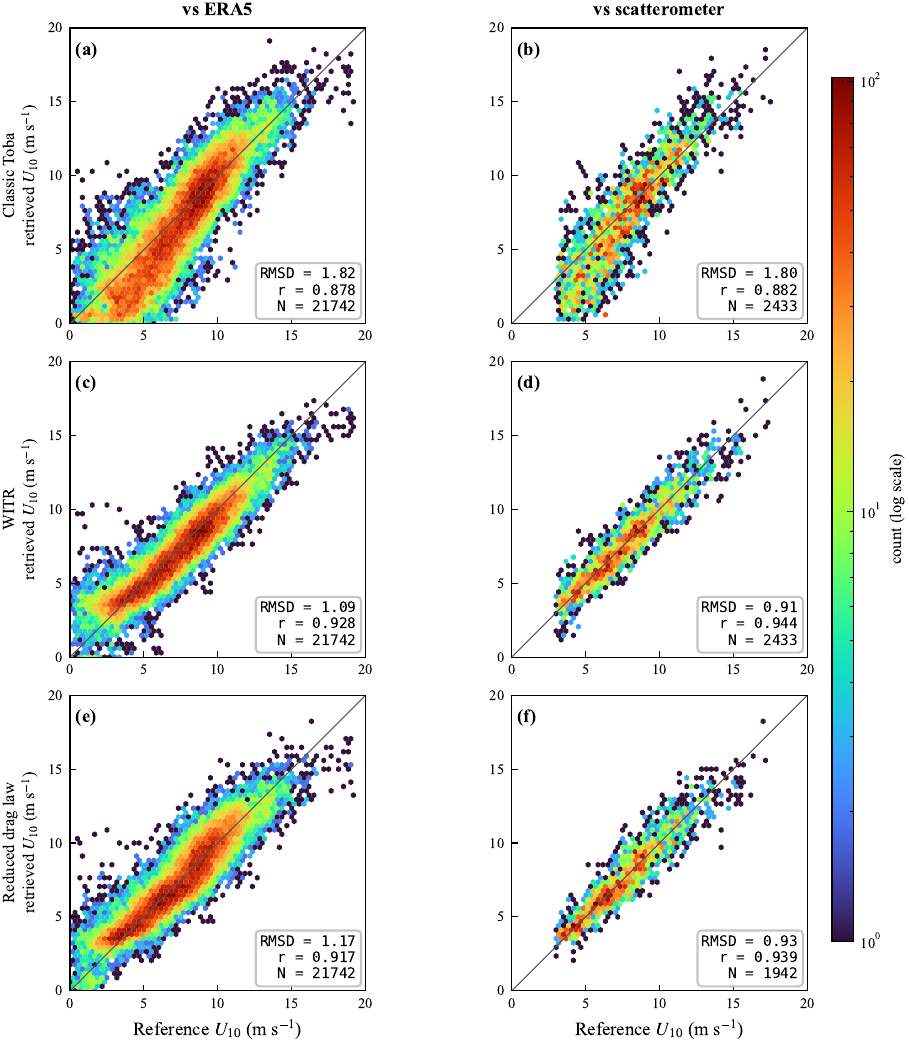}}
  \caption{Wind-speed validation for the three principal retrievals against
  both references, as two-dimensional density histograms (hexagonal bins, log colour scale) with
  the 1:1 line and statistics. Rows: classic Toba (a,~b), WITR (c,~d),
  reduced drag law (e,~f). Columns: retrieved versus ERA5 (left) and
  versus the pooled scatterometer winds (right). RMSD (m~s$^{-1}$): Toba
  1.82/1.80, WITR 1.09/0.91, reduced drag law 1.17/0.93 (ERA5/scat).}
  \label{fig:speed_scatter}
\end{figure*}

Table~\ref{tab:speed_val} shows a clear progression from the traditional
single-band Toba inversion (1.80~m~s$^{-1}$ vs.\ pooled scat) through the
analytical fits to WITR (0.91~m~s$^{-1}$ scat, 1.09~m~s$^{-1}$ ERA5). The
larger
difference against ERA5 does not indicate a worse retrieval there: ERA5
under-represents mesoscale wind variability, so the scatterometer is the
better representation of the wind actually experienced by the buoy, and
the ERA5 comparison absorbs ERA5's own smoothing and biases
\citep{BelmonteStoffelen2019}. The classic Toba
inversion is the noisiest throughout the range
(Fig.~\ref{fig:speed_scatter}a,b), consistent
with the $\sim$2~m~s$^{-1}$ reported for the single-band Sofar method
\citep{Voermans2020,BeckmanLong2022}. WITR was developed as a Ridge+GBM
model on the full 52-feature pool ($\sim$1.00~m~s$^{-1}$ leave-one-buoy-out)
then distilled by backward elimination to the nine acceleration-spectrum
features used throughout ($\sim$0.91~m~s$^{-1}$;
Fig.~\ref{fig:speed_scatter}c,d), which are available for all data tiers.
Its sigmoid-gated architecture holds strong-wind ($>$12~m~s$^{-1}$) RMSD at
1.19~m~s$^{-1}$, where the spectrum-only spectral law degrades and the
single-band inversion is noisier throughout
(Table~\ref{tab:speed_regime}); across sources the C-band ASCAT instruments
give slightly lower RMSD (0.86--0.87~m~s$^{-1}$) than the coarser
25-km-grid Ku-band HY-2 (0.96--0.98~m~s$^{-1}$; per-satellite statistics in Table~S1 of the online supplemental material); the C- and
Ku-band samples also differ through rain screening, which removes a larger
share of Ku-band cells \citep{VogelzangStoffelen2022}. The
closed-form reduced drag law [Eq.~(\ref{eq:reduced_drag})] reaches
0.93~m~s$^{-1}$ vs.\ scat (0.94 leave-one-buoy-out, suggesting limited buoy-level optimism) with
four dimensionless groups (Fig.~\ref{fig:speed_scatter}e,f); its high-wind
term holds strong-wind RMSD at 1.28~m~s$^{-1}$---well below the spectrum-only
spectral law's 1.87---so only WITR's gradient-boosted residual correction
still separates the two.

\clearpage
\clearpage

The wind-speed dependence of the bias is assessed in 2~m~s$^{-1}$ bins of
the \emph{mean} of the retrieved and reference winds; binning the
difference by either wind alone creates artificial speed-dependent biases
at both ends of the range through regression to the mean
\citep{Stoffelen1998,Ni2022}---binning by the scatterometer wind alone
would give $+0.2$~m~s$^{-1}$ in the lowest bin, vanishing under symmetric
binning. Binned this way, WITR is conditionally
unbiased against the pooled scatterometer
($|\mathrm{bias}| \leq 0.12$~m~s$^{-1}$) across the entire sampled
2--16~m~s$^{-1}$ range. The collocated sample nevertheless thins rapidly
toward both extremes---calm winds below 3~m~s$^{-1}$ are additionally
subject to the scatterometer quality control---so the per-bin statistics
outside 4--14~m~s$^{-1}$ rest on modest sample sizes and we refrain from
over-interpreting them.

\subsection{Analytical equation performance}

The reduced drag law [Eq.~(\ref{eq:reduced_drag})] provides an
complementary analytical assessment of the wind retrieval, built on a
lower-dimensional feature set that also includes buoy-motion inputs (four dimensionless groups derived from
$\beta_4$ in three bands, $\sigma_{a_z}$, and pitch RMS) compared to
the nine acceleration-band features of WITR. Because the closed form
requires the IMU motion inputs $\sigma_{a_z}$ and pitch RMS, it can be
evaluated only on the four recovered buoys (4\,437
collocations, 67\% of the L2B set); the 22 historical satellite-transmitted
buoys record spectra but not motion data. On that subset the reduced drag
law achieves RMSD = 0.97~m~s$^{-1}$ (Table~S1), about
0.05~m~s$^{-1}$ above WITR. It is closest to WITR for the C-band ASCAT
instruments (0.90--0.93~m~s$^{-1}$) and higher for the Ku-band HY-2 instruments
(1.00~m~s$^{-1}$), consistent with the larger representativeness
noise of the coarser 25-km Ku-band product.

Per-regime, the reduced drag law tracks WITR within
$\sim$0.1~m~s$^{-1}$ across most of the wind range, including strong winds
where its high-wind drag-reduction term holds the $>$12~m~s$^{-1}$ RMSD at
1.28~m~s$^{-1}$ (versus WITR's 1.19); only WITR's nonlinear
gradient-boosted correction above the sigmoid threshold separates them.

The two retrievals are not independent: on the matched recovered-buoy
subset ($N = 1\,942$) their residuals against the scatterometer correlate at
$\rho = 0.81$, so averaging them improves the RMSD only marginally (0.86
versus 0.88~m~s$^{-1}$ for WITR alone; details in section~S3 of the
online supplemental material).

\subsection{Direction validation}\label{sec:dir_val}

Table~\ref{tab:dir_val} validates wind direction against independent
scatterometer collocations using the mean absolute angular difference
(MAAD, the mean of the wrapped angular difference between the two
systems); the GNSS drift bearing (over a
$\pm$1.5~h centered window) and ERA5 are included for comparison.

\begin{table}[t]
\caption{Wind direction validation. MAAD and bias in degrees. The
  $\geq$4~m~s$^{-1}$ subset is shown separately because angular direction
  becomes intrinsically ill-defined as wind speed approaches zero; the
  vector-component statistics retain the low-speed comparisons. OTC25\_20 contributes the remaining 8 of the 521
  ASCAT~L2 collocations---too few for a separate per-buoy row.
  $^\ddagger$Drift bearing over a $\pm$1.5-h window, read as wind
  direction via drift-toward~$+180^{\circ}$.}\label{tab:dir_val}
\begin{center}
\small
\setlength{\tabcolsep}{4pt}
\resizebox{\columnwidth}{!}{%
\begin{tabular}{llrrr}
\hline\hline
Method & Buoy & $N$ & MAAD ($^{\circ}$) & Bias ($^{\circ}$) \\
\hline
\multicolumn{5}{c}{\textit{IMU $D_p$ (0.6--0.9~Hz) vs.\ ASCAT L2}} \\
\hline
 & OTC25\_04  & 250 & 10.5 & $+$2.5 \\
 & EXPLOI\_06 & 155 & 5.8 & $-$0.5 \\
 & OTC25\_16  & 108 & 9.3 & $+$7.4 \\
 & All L2     & 521 & 9.4 & $+$3.1 \\
\hline
\multicolumn{5}{c}{\textit{Reference baselines on the same 521 collocations}} \\
\hline
ERA5        & All  & 521 & 10.8 & $+$1.2 \\
GNSS drift bearing$^\ddagger$ & All & 521 & 53.0 & $+$16.2 \\
\hline
\multicolumn{5}{c}{\textit{IMU $D_p$ (0.6--0.9~Hz) vs.\ multi-satellite L2B}} \\
\hline
 & All L2B                       & 4\,437 & 9.6 & $+$2.4 \\
 & All L2B ($\geq$4~m~s$^{-1}$)  & 4\,085 & 8.8 & $+$2.2 \\
\hline\hline
\end{tabular}}
\end{center}
\end{table}

\begin{figure*}[t]
  \centerline{\includegraphics[width=\textwidth]{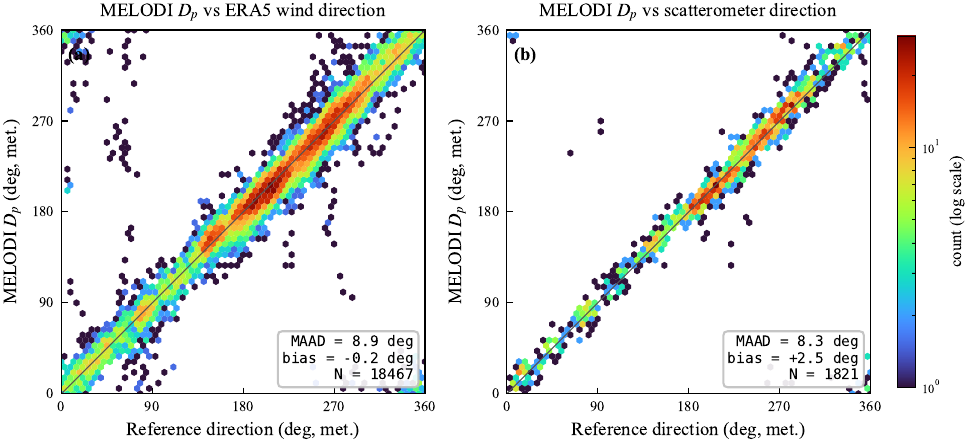}}
  \caption{MELODI wind-sea $D_p$ (0.6--0.9~Hz, filtered) versus (a)~ERA5 direction (MAAD 8.9$^{\circ}$) and (b)~the scatterometer reference (MAAD 8.3$^{\circ}$), for winds above 4~m~s$^{-1}$. The 1:1 line and circular statistics are shown.}
  \label{fig:dir_scatter}
\end{figure*}

The IMU $D_p$ achieves a MAAD of 9.4$^{\circ}$ against 521 ASCAT L2
collocations (best at the trade-wind EXPLOI\_06, 5.8$^{\circ}$), versus 53$^{\circ}$ for the GNSS drift bearing
(Table~\ref{tab:dir_val}), which is therefore not a useful
wind-direction estimate. The difference is wind-speed dependent, from above
25$^{\circ}$ in the lightest ERA5-speed bins (near 3~m~s$^{-1}$, where
the matched scatterometer subset of section~\ref{sec:dir_val} gives
13$^{\circ}$ in mean-wind bins) to $\sim$7$^{\circ}$ at
13~m~s$^{-1}$: stronger winds raise the wind-sea coherence $r_1$ and
sharpen the directional moments. On the larger multi-satellite L2B set the
MAAD is 9.6$^{\circ}$, falling to 8.8$^{\circ}$ above 4~m~s$^{-1}$. For
winds above 4~m~s$^{-1}$ the MELODI $D_p$ reaches 8.9$^{\circ}$ against
ERA5 and 8.3$^{\circ}$ against the tight-QC scatterometer winds
(Fig.~\ref{fig:dir_scatter}).

\paragraph*{Vector-component view.}
Because speed statistics are skewed and direction is ill-defined at low
wind, the wind-vector components provide the cleaner low-wind metric
\citep{VogelzangStoffelen2022}: on the matched recovered-buoy subset
($N = 1\,942$) the zonal and meridional component differences between the
MELODI vector (WITR speed, IMU $D_p$) and the scatterometer are 1.36 and
1.31~m~s$^{-1}$ RMSD (biases $-0.0$ and $-0.2$). Unlike the speed and
direction metrics, the components show no low-wind anomaly---they are
smallest at low wind ($\approx$0.9~m~s$^{-1}$ at 3--4~m~s$^{-1}$, where
the direction MAAD is 13$^{\circ}$, and 10$^{\circ}$ at
4--5~m~s$^{-1}$) and grow with wind speed as angular
differences project onto increasingly long vectors. The direction
retrieval is therefore reported at low winds rather than withheld.

\paragraph*{Band sensitivity.}
The choice of the 0.60--0.90~Hz wind-sea band is justified quantitatively
in Table~\ref{tab:band_sens}, which re-extracts $D_p$ in three candidate
bands through the identical pipeline on the same recovered-buoy sessions.
Moving from the conventional 0.35--0.60~Hz band to 0.60--0.90~Hz lowers the
MAAD from 9.4$^{\circ}$ to 8.9$^{\circ}$ against ERA5 and from
9.3$^{\circ}$ to 8.3$^{\circ}$ against the scatterometer; pushing higher
(0.70--1.00~Hz) yields no further gain. Notably, the conventional band
carries the \emph{highest} directional coherence (median $r_1 = 0.56$
versus $0.43$) yet the largest error---the signature of long-period swell
leaking coherent but mis-aligned energy into the directional moments,
precisely the contamination the higher band avoids. The aggregate
improvement is modest because it is concentrated in mixed seas where swell
and wind-sea coexist: at the swell-exposed OTC25\_16 buoy the ERA5 MAAD falls
from 10.2$^{\circ}$ (0.35--0.60~Hz) to 8.2$^{\circ}$ (0.60--0.90~Hz),
whereas the trade-wind EXPLOI\_06 record, dominated by a clean wind-sea, is
essentially band-independent (7.2$^{\circ}$ versus 7.3$^{\circ}$).

Table~S2 in the online supplemental material repeats the same
three-band comparison separately by inverse wave age $U_{10}/c_p$,
wind-sea energy fraction, and wind speed. The chosen band performs best precisely in the
difficult regimes: in light winds it is essentially tied with the
higher-frequency band and clearly outperforms the conventional one, and
in swell-influenced seas it is the best of the three. The conventional
band recovers 0.5--0.9$^{\circ}$ in mature seas at fresh-to-strong
winds, and actively growing seas ($U_{10}/c_p > 1.2$, 0.7\% of sessions)
degrade all bands alike. Where the low- and high-band directions oppose each other
($\approx$1\% of sessions), no fixed band performs well---these are the
swell--wind-sea transitions already captured by the $r_1$ flag. A
regime-adaptive selection could therefore recover about 1$^{\circ}$ at
most, and only where errors are already smallest; we retain the fixed
0.60--0.90~Hz band for robustness and onboard simplicity.

\begin{table}[t]
\caption{Wind-direction band sensitivity. Circular MAAD (winds
  $\geq$4~m~s$^{-1}$) of the filtered $D_p$ against ERA5 and the tight-QC
  scatterometer reference, the fraction of low-coherence ($r_1 < 0.2$)
  sessions, and the median $r_1$, for three candidate wind-sea bands. All
  bands, and the main validation (Table~\ref{tab:dir_val},
  Fig.~\ref{fig:dir_scatter}), are computed through the identical pipeline on
  the same 21\,742 recovered-buoy sessions, so the chosen-band values match
  the main validation exactly and the columns are directly comparable.}%
  \label{tab:band_sens}
\begin{center}
\small
\setlength{\tabcolsep}{5pt}
\resizebox{\columnwidth}{!}{%
\begin{tabular}{lrrrr}
\hline\hline
Band (Hz) & MAAD/ERA5 & MAAD/scat & flagged & median \\
          & ($^{\circ}$) & ($^{\circ}$) & (\%) & $r_1$ \\
\hline
0.35--0.60 (conventional) & 9.4 & 9.3 & 1.0 & 0.56 \\
0.60--0.90 (\textbf{chosen}) & 8.9 & 8.3 & 1.7 & 0.43 \\
0.70--1.00 (high)            & 9.1 & 8.5 & 1.8 & 0.39 \\
\hline\hline
\end{tabular}}
\end{center}
\end{table}

\subsection{Wind-vector time series}\label{sec:vector_ts}

\begin{figure*}[t]
  \centerline{\includegraphics[width=\textwidth]{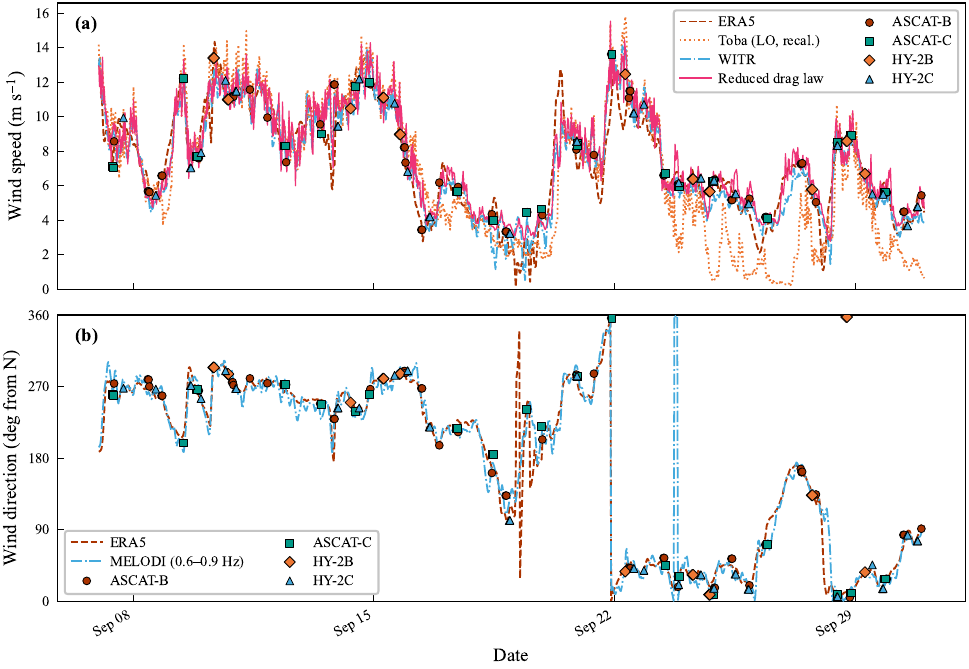}}
  \caption{Wind-vector time series at OTC25\_04 (moderate-to-strong-wind period,
    September~2025; ERA5 peak $\sim$15~m~s$^{-1}$).
    Top: wind speed from ERA5, the classic Toba inversion, WITR, the
    reduced drag law, and the collocated scatterometer overpasses.
    Bottom: wind direction from ERA5, the filtered MELODI $D_p$, and the
    scatterometer passes.}
  \label{fig:vector_ts}
\end{figure*}

Figure~\ref{fig:vector_ts} assembles the speed and direction retrievals
into a single wind-vector time series for the OTC25\_04 buoy during a
moderate-to-strong-wind period in September~2025 (ERA5 peak $\sim$15~m~s$^{-1}$). The
speed panel overlays the ERA5 reference, the classic Toba inversion, the
WITR production model, the reduced drag law, and the collocated
scatterometer overpasses: WITR tracks ERA5 and the scatterometer passes
closely across the full range of wind variations, the reduced drag law
follows within $\sim$1~m~s$^{-1}$, and the classic Toba inversion is
noisier and tends to underestimate the stronger winds. The direction panel
overlays ERA5, the filtered MELODI $D_p$, and the scatterometer passes,
tracking the direction shifts closely.

\subsection{Bias structure across the wind range}\label{sec:error_summary}

\begin{figure*}[t]
  \centerline{\includegraphics[width=\textwidth]{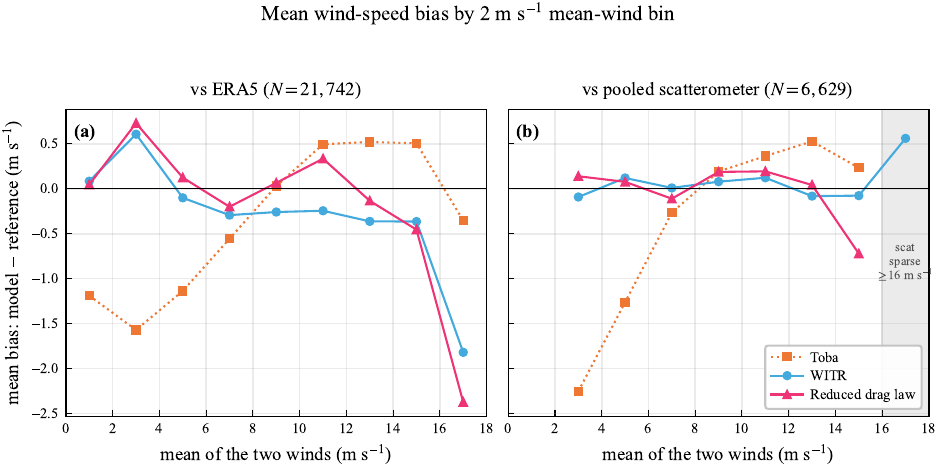}}
  \caption{Mean wind-speed bias (model $-$ reference) in 2~m~s$^{-1}$ bins
    of the mean of the two collocated winds \citep{Stoffelen1998}, for
    Toba, WITR, and the reduced drag law, against (a)~ERA5 and (b)~the
    pooled scatterometer, on a common wind axis. The shaded band in (b)
    marks $\geq$16~m~s$^{-1}$, where the
    scatterometer data set retains too few collocations to constrain the
    bias.}
  \label{fig:bias_bins}
\end{figure*}

Figure~\ref{fig:bias_bins} resolves the wind-speed-dependent bias of the
three principal models in 2~m~s$^{-1}$ bins of the mean of the two
collocated winds, for both references---more diagnostic than a single
pooled RMSD. Against the scatterometer, WITR and the reduced drag law are
near-unbiased over the well-populated 4--14~m~s$^{-1}$ range; the reduced
drag law develops a modest underestimate ($-0.7$~m~s$^{-1}$) at
14--16~m~s$^{-1}$, and the single-band inversion, even after its global
rescaling, underestimates light winds by 1--2~m~s$^{-1}$ and carries 1.5--2
times the per-bin spread of the learned retrievals. Against ERA5 the learned and
analytical retrievals stay within $\pm$0.5~m~s$^{-1}$ up to
$\sim$15~m~s$^{-1}$ and drift negative only in the sparsely populated
highest bins; this excursion is absent against the scatterometer and
partly reflects ERA5's own limitations at the extremes
\citep{BelmonteStoffelen2019}. Above
16~m~s$^{-1}$ the collocation sample is too thin to constrain the bias
with either reference (section~\ref{sec:limitations}).

\section{Operational handling of the retrievals}\label{sec:product}

To show that the retrievals scale to a multi-buoy fleet, this section
demonstrates how the WITR wind speed and the IMU-based direction are
combined, quality-flagged, and assembled into a single wind-vector dataset
(written as a NetCDF file), and reports the resulting multi-buoy
performance. On the newest MELODI drifters the wind retrieval also runs
onboard: the band-integrated directional moments and the session wind
vector are computed on the buoy and transmitted with the spectral
features, so unrecovered drifters report both components; the centered
temporal filtering and final quality control remain shore-side.

\subsection{Dataset description}

The assembled wind-vector dataset contains the 29\,775 sessions of
Table~\ref{tab:datasets} (28 buoys, March 2025--March 2026, four ocean
basins). Each session carries the WITR wind speed, wind direction (Tier~B
only, 73\% of sessions), friction velocity $u_*$, position, buoy
identifier, processing tier, and separate speed and direction quality
flags. A speed value is available for every session; 1.9\% are flagged as
low confidence because of anomalous or incomplete spectral inputs, and
winds above 17~m~s$^{-1}$ would be flagged as extrapolative (none occur
in the present record). For Tier~B
sessions the direction flag identifies low wind-sea coherence
($r_1 < 0.2$; 1.7\% of Tier~B sessions); Tier~A sessions carry no
direction estimate.

\subsection{Multi-buoy performance}\label{sec:multibuoy}

The WITR model is validated per buoy by leave-one-buoy-out
cross-validation against scatterometer winds (the per-reference bias
structure is shown in Fig.~\ref{fig:bias_bins}).
Across the 17 tight-QC buoys with at least 10 collocations, the per-buoy
RMSD for wind speed ranges from 0.70 to 1.28~m~s$^{-1}$ (median
0.96~m~s$^{-1}$). The four Tier~B buoys with the most extensive
collocation records---OTC25\_04, EXPLOI\_06, OTC25\_16, and
OTC25\_20---retrieve at RMSD near or below 1.0~m~s$^{-1}$ (0.92, 0.75,
0.96, and 1.01~m~s$^{-1}$ respectively, with EXPLOI\_06 in the Indian
Ocean trade-wind belt the most accurate), demonstrating that the WITR
model generalizes across ocean basins. The largest per-buoy RMSDs
($>$1.2~m~s$^{-1}$) occur at historical satellite-transmitted buoys with
only a handful of collocations, where the small sample size dominates
the RMSD.

For Tier~A buoys (satellite-transmitted spectra), the WITR model is
applied using the same 9 acceleration-spectrum features extracted from
the onboard-computed $S_{\mathrm{acc}}(f)$. Evaluated against ERA5, the
median per-buoy RMSD across the 24 Tier~A buoys is 1.4~m~s$^{-1}$---above
the sub-metre errors achieved against scatterometers on the recovered
Tier~B buoys, reflecting the different onboard spectral processing and
compression of the satellite-transmitted data (128 bins, no merging)
compared with the full Welch estimation from recovered data, and ERA5's
larger representativeness
error relative to the scatterometer collocation.

\section{Discussion}\label{sec:discussion}

\subsection{Comparison with established anemometer networks}

Table~\ref{tab:comparison} places the MELODI results in the context
of published moored-buoy validation statistics. The global
NDBC--TAO--PIRATA--RAMA anemometer network achieves RMSD of
1.11--1.24~m~s$^{-1}$ against ASCAT across 206~moored buoys and more
than $1.1 \times 10^5$~collocations \citep{YangEA2019}. The official
KNMI/OSI~SAF ASCAT validation reports an equivalent scatter of
$\sim$1.05~m~s$^{-1}$ \citep{VerhoefStoffelen2013}. Against the HY-2
series (Ku-band), moored buoys achieve 0.95--1.07~m~s$^{-1}$
\citep{ZhaoEA2021,ZhaoEA2023}.

The MELODI WITR model, despite lacking a dedicated wind sensor,
achieves a pooled-scatterometer RMSD of 0.91~m~s$^{-1}$ by
leave-one-buoy-out cross-validation (0.93 on the full L2B set, in-sample for
the recovered buoys). This places its scatterometer difference at
the lower end of the published moored-buoy--scatterometer range and on a par
with the best HY-2 configurations. We stress that this is a comparison of
\emph{differences against the same scatterometer reference}, not a claim of
equivalence to a calibrated mast anemometer: the moored figures are
themselves buoy-minus-scatterometer scatter, and both terms carry
representativeness and reference error. Cross-study comparisons of such
differences carry limited precision in any case: different collocation
data sets sample different weather and apply different quality control,
so differences of a few tenths of a m~s$^{-1}$ between studies are not
significant \citep{Wright2021}. The reduced drag law
(0.97~m~s$^{-1}$ on the recovered-buoy subset) matches the KNMI/OSI~SAF
scatterometer-difference benchmark ($\sim$1.05~m~s$^{-1}$) from four
dimensionless groups and no machine-learning components. The combined
wind-vector difference, the magnitude of
$|\mathbf{U}_{\mathrm{WITR}} - \mathbf{U}_{\mathrm{scat}}|$, has a
median of 1.2~m~s$^{-1}$ across all L2B collocations with valid
direction, likewise comparable to the vector differences reported for
moored anemometer buoys against scatterometers (1.1--1.5~m~s$^{-1}$;
\citealp{YangEA2019}).

\begin{table*}[t]
\caption{Wind speed RMSD (m~s$^{-1}$) against scatterometer references:
  MELODI models compared with published moored-buoy benchmarks and
  triple collocation estimates.}\label{tab:comparison}
\begin{center}
\begin{tabular}{lcccl}
\hline\hline
System & RMSD & Bias & $N$ & Reference \\
\hline
\multicolumn{5}{c}{\textit{MELODI (this study)}} \\
WITR (LOBO)             & 0.91  & $-$0.01  & 2\,433   & 5~products$^*$ \\
WITR (full L2B)         & 0.93  & $+$0.06  & 6\,629   & 4~satellites \\
Reduced drag law        & 0.97  & $+$0.05  & 4\,437   & 4~satellites \\
Extended spectral law   & 0.98  & $-$0.12  & 4\,437   & 4~satellites \\
Toba (single-band)      & 1.75  & $-$0.41  & 4\,437   & 4~satellites \\
\hline
\multicolumn{5}{c}{\textit{Moored anemometer buoys (literature)}} \\
NDBC + TAO (206~buoys)  & 1.11--1.24 & ---  & $>$110\,000 & \citet{YangEA2019} \\
KNMI/OSI SAF            & $\sim$1.05 & ---  & $\sim$2\,300 & \citet{VerhoefStoffelen2013} \\
HY-2B (Ku-band)         & 0.95--1.07 & ---  & ---       & \citet{ZhaoEA2021} \\
HY-2C/D (Ku-band)       & 0.78--1.03 & ---  & ---       & \citet{ZhaoEA2023} \\
\hline
\multicolumn{5}{c}{\textit{Drifter wind (literature)}} \\
Spotter fleet (global)  & $\sim$1.0  & ---  & $>$20\,000 & \citet{Dorsay2023} \\
Spotter (Toba)          & $\sim$2.0  & ---  & ---       & \citet{Voermans2020} \\
\hline
\multicolumn{5}{c}{\textit{Triple collocation error SD}} \\
ASCAT (C-band)          & $\sim$0.7  & ---  & ---       & \citet{VogelzangStoffelen2021} \\
ERA5                    & 0.8--1.1   & ---  & ---       & \citet{VogelzangStoffelen2021} \\
\hline\hline
\end{tabular}
\end{center}
{\footnotesize $^*$The leave-one-buoy-out (LOBO) validation pools the
  tight-QC data set of five scatterometer products spanning the same four
  satellites: MetOp-B contributes both the Ocean and Sea Ice Satellite
  Application Facility (OSI SAF) OSI-104 coastal product and
  the standard 12.5-km product (both Level-2B), plus MetOp-C and
  HY-2B/C. The full-L2B and analytical
  evaluations use only the four-satellite L2B products. The motion-based
  reduced drag law is evaluable on the 4\,437 recovered-buoy collocations
  that carry the required IMU inputs. For a like-for-like comparison with the
  reduced drag law, the spectrum-only extended spectral law is also reported
  on this same 4\,437-collocation recovered-buoy subset; being spectrum-only,
  it is in principle applicable to all 6\,629 L2B collocations and to the
  Tier~A satellite-transmitted spectra (where it is evaluated against ERA5 in
  section~\ref{sec:results}, median per-buoy RMSD 1.4~m~s$^{-1}$), but the
  matching L2B-collocated $\beta_4$ inputs were assembled only for the
  recovered buoys.}
\end{table*}

\subsection{Co-deployment with a Sofar Spotter}\label{sec:spotter}

The Sofar Spotter is the established commercial drifter that operationally
retrieves wind from the wave spectrum via the single-band inversion of
\citet{Voermans2020}, and is the natural benchmark for our retrieval. During
the OTC25 campaign a Spotter (SPOT-32265C) drifted alongside the recovered
MELODI buoy OTC25\_16 for 14~days (14--28~May~2025, NE~Atlantic,
$\sim$55$^\circ$N 17$^\circ$W; within $<$30~km for the first week). We
compare both against the same ERA5 reference on the 665 collocated hourly
samples (Table~S4 in the online supplemental material), treating the
Spotter as a peer instrument, not ground truth.

\begin{figure*}[t]
  \centerline{\includegraphics[width=\textwidth]{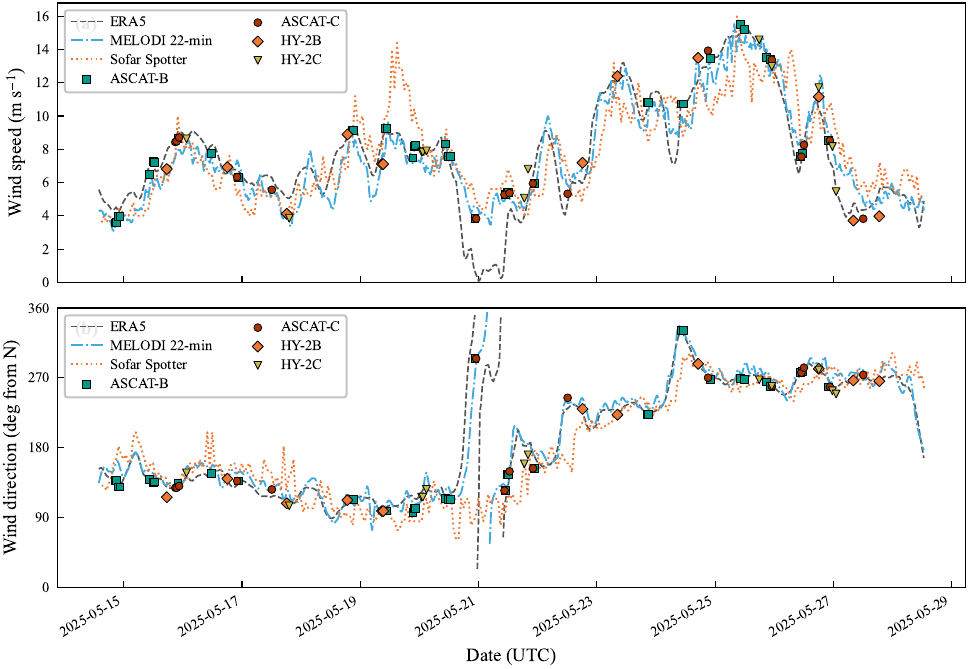}}
  \caption{MELODI versus Sofar Spotter over their 14-day co-deployment (OTC25\_16 $\leftrightarrow$ SPOT-32265C, 14--28~May~2025, NE~Atlantic). The MELODI product is compared against the Sofar Spotter, ERA5, and the collocated scatterometer overpasses for (top) wind speed and (bottom) wind direction. The Spotter is noticeably noisier at low wind.}
  \label{fig:sofar_ts}
\end{figure*}

Against ERA5, the MELODI series is both the tighter (RMSD
1.34 vs 2.10~m~s$^{-1}$; the Spotter consistent with the $\sim$2~m~s$^{-1}$ of
\citet{Voermans2020}) and essentially unbiased in this window
($-$0.04 vs $+$0.14~m~s$^{-1}$). MELODI direction is substantially better
(16.5$^\circ$, $r_{\mathrm{circ}}=0.98$ vs 40.0$^\circ$, 0.83), the gap
concentrated at low wind: below 5~m~s$^{-1}$ the Spotter degrades to
81$^\circ$ RMSD---direction being poorly defined when the wind-sea is
weak---whereas the despiked MELODI estimate holds at 29$^\circ$; above
8~m~s$^{-1}$ both are accurate (MELODI 10$^\circ$, Spotter
16$^\circ$). The co-deployment
time series (Fig.~\ref{fig:sofar_ts}) shows the Spotter visibly noisier at
low wind.

The difference is methodological: the Spotter reads a single equilibrium
level and the $(a_1,b_1)$ at that band, whereas MELODI combines a
multi-band, scatterometer-trained speed regression with high-frequency
directional moments and circular despiking. The comparison is necessarily
illustrative (a single 14-day window, ERA5 as the common reference, and
dependent on the Spotter processing version and deployment configuration),
but within those limits MELODI had the smaller differences from ERA5 in
both components, the difference being largest at low wind. The signature is not unique to this window: it is consistent
with the yearlong moored-Spotter statistics of \citet{BeckmanLong2022}
(direction $95.7^\circ$ RMSD; speed $2.1$~m~s$^{-1}$ with a low bias),
suggesting the low-wind limitation is largely methodological rather than
site-specific.

\subsection{Accuracy floor and representativeness}\label{sec:accfloor}

The achievable accuracy for any buoy-based wind estimate validated
against scatterometers is bounded by the scatterometer's own measurement
error. Quintuple collocation analyses
\citep{VogelzangStoffelen2021,VogelzangStoffelen2022} estimate wind
component error standard deviations below 0.5~m~s$^{-1}$ for the C-band
ASCATs and 0.6--0.7~m~s$^{-1}$ for the Ku-band scatterometers, with
numerical-weather-prediction model values of 0.85--1.0~m~s$^{-1}$ and buoy terms of
0.9--1.1~m~s$^{-1}$ on the scatterometer scale (including
representativeness). The buoy leg carries the largest collocation-scale variance among the
three systems, dominated by spatial representativeness rather than
anemometer noise: a point measurement
is compared with a satellite cell of 12.5--25-km grid spacing, and
subgrid-scale turbulent fluctuations contribute $\sim$0.5~m~s$^{-1}$ to
the mismatch \citep{Wright2021}.

Triple collocation of our own triplets (MELODI, scatterometer, ERA5;
$N = 2\,433$; section~S4 of the online supplemental material) attributes
error standard deviations of $\approx$0.9~m~s$^{-1}$ to MELODI and
0.71--0.81~m~s$^{-1}$ to ERA5, consistent with the pairwise RMSDs and the
published estimates above, and is consistent with the
$\approx$0.9~m~s$^{-1}$ pairwise difference being dominated by the MELODI
error and representativeness rather than by scatterometer noise. The scatterometer
term of this decomposition (0.27~m~s$^{-1}$) is reported in the
supplement only as a diagnostic of the failed independence assumptions
(WITR is trained against the scatterometer, and ERA5 assimilates it); the
independent values of \citet{VogelzangStoffelen2022} remain the reference
for the scatterometer error.

The MELODI LOBO RMSD of 0.91~m~s$^{-1}$ thus combines (i)~the model
retrieval error, (ii)~the scatterometer measurement error, and
(iii)~spatial and temporal representativeness noise; the retrieval
contribution is therefore smaller than the pairwise RMSD, but the
published component-error values cannot be subtracted directly from a
scalar wind-speed difference, so we do not quote a numerical
decomposition. Further improvement beyond $\sim$0.9~m~s$^{-1}$ RMSD would therefore
require either higher-resolution scatterometer references or a tighter
collocation window. No dependence on temporal offset was detectable
within the adopted collocation window (RMSD 0.71--0.81~m~s$^{-1}$ across
$|\Delta t|$ bins; rank correlation $r = 0.01$, $N = 1{,}942$), so the
residual difference is limited by the scatterometer measurement and
spatial representativeness rather than temporal mismatch.

A further consideration concerns what wind is being compared. Scatterometers
are calibrated to stress-equivalent winds ($U_{10S}$; \citealp{deKloe2017}):
the backscatter responds to the centimetre-scale roughness set by the
surface stress, independent of atmospheric stability and air density. The
drifter retrieval is of the same kind---it senses the sea state, not the
atmosphere---so it is naturally closer to a stress-equivalent than to an
anemometer wind, and part of any difference against anemometer-based
references (locally up to $\pm$0.5~m~s$^{-1}$) would reflect stability
and air-density effects rather than error of either system.

The scatterometer reference also defines the scale of the product. WITR
is trained on scatterometer winds because they are at present the only
source of collocated wind vectors numerous enough for training, and the
best-characterized open-ocean wind reference, with a long record of
calibration and validation against moored buoys and numerical weather
prediction \citep{Stoffelen1998,VerhoefStoffelen2013,VogelzangStoffelen2022}.
The retrieved speed is therefore expressed on the scatterometer scale,
whereas the direction uses no reference wind; for the scatterometers
themselves, the drifters test consistency across missions and regions
rather than the absolute scale. The method is not tied to this reference:
the same features and regression can be recalibrated against anemometer
winds as collocated measurements accumulate. On this scale, a drifter
network extends calibration and validation to places and times the
scatterometers do not sample, serving SAR wind fields \citep{Mouche2019},
altimeter and near-nadir wind and sea-state products (including CFOSAT
SWIM and SWOT; \citealp{RibalYoung2019,Hauser2021,Morrow2019}), and
atmospheric and wave models that need continuous point time series.

\subsection{Physical interpretability: analytical equations}

A notable result of this study is that the symbolic regression discovers
a closed-form dimensionless drag law [Eq.~(\ref{eq:reduced_drag});
section~\ref{sec:analytical}] that reaches an RMSD of 0.93~m~s$^{-1}$
against the scatterometer winds from four dimensionless groups. Its
structure---a canonical logarithmic base drag, a smooth-to-rough viscous
correction, a wave-age modulation, and an empirical high-wind
correction---contains terms analogous to those of bulk-flux formulations
\citep{Fairall2003}, evaluated from buoy IMU inputs alone at inference
time, in a form
executable onboard in $\sim$20 floating-point operations. The physical
reading of each term, and the caveat that the high-wind term turns the
implied drag coefficient over well below the observed $C_d$ saturation
\citep{Powell2003,Donelan2004,Edson2013,CurcicHaus2020}, are given in
section~\ref{sec:analytical} and expanded in section~S5 of the online
supplemental material.

\subsection{Limitations and future refinements}\label{sec:limitations}

Several limitations bound the present results. (1)~The inversion relies on
the equilibrium-range assumption, which breaks down under rapid wind
changes (timescales $<\sim$30~min), strong swell--wind-sea coupling, and
very light winds ($<$3~m~s$^{-1}$); the calm-wind regime accordingly
carries the largest relative error (0--5~m~s$^{-1}$ tight-QC regime,
RMSD 0.88~m~s$^{-1}$, $\sim$30\% of the bin mean;
Table~\ref{tab:speed_regime}). This floor is intrinsic to any wave-based proxy: at low winds
momentum exchange shifts from form drag to viscous drag over the smooth
surface \citep{KudryavtsevMakin2001}, and the reference-trained
regression reduces but cannot eliminate the effect. The same mechanism
has been invoked for the single-band Spotter's low-wind underestimation
\citep{Voermans2020,BeckmanLong2022}. (2)~The WITR GBM residual correction
limits full physical transparency, although the analytical drag law closes
most of the gap (0.97 versus 0.93~m~s$^{-1}$ on the L2B set). (3)~Wind
direction requires directional wave moments, derived from GNSS or IMU
measurements; a transmitted heave spectrum alone yields only wind speed.
Retrieving the full wind vector from unrecovered drifters therefore
requires either transmitting this additional spectral information or
computing it onboard, as the newest MELODI firmware does. (4)~The inversion reads sea state through the
buoy's own motion, so hull response and platform dynamics enter the
high-frequency spectrum; band selection and filtering mitigate but cannot
remove this, and it is expected to differ across hull designs.

Most importantly, the calibration samples a limited envelope. The reference
data thin out steeply with wind speed---the tight-QC scatterometer set
used for the speed validation holds only a few dozen collocations above
14~m~s$^{-1}$ and none above 17.4~m~s$^{-1}$, and the full L2B set reaches
21~m~s$^{-1}$---while ERA5, which reaches 19~m~s$^{-1}$ in our
collocations, is itself least reliable at strong winds. Strong-wind behaviour is
therefore an \emph{extrapolation} rather than a validated result;
accordingly, winds above 17~m~s$^{-1}$---the highest wind the tight-QC
scatterometer collocations reach---are flagged as low-trust extrapolation
in the product. Targeted strong-wind collocations would directly
constrain the high-wind drag-reduction term [$-0.00243\,\mathrm{Re}_*$ in
Eq.~(\ref{eq:reduced_drag})], presently fit on this sparse tail.
Under-represented mixed and cross-sea states can likewise place swell energy
in the wind-sea bands and bias both the speed level and the directional
moments. The practical remedy is wide deployment: buoy--scatterometer
collocation archives demonstrate that satellite validation accumulates
strong-wind matchups far faster than dedicated in situ campaigns
(350\,000 ASCAT--buoy pairs reaching 25~m~s$^{-1}$ in the C-band High
and Extreme-Force Speeds study, CHEFS, of the European Organisation for
the Exploitation of Meteorological Satellites, EUMETSAT;
\citealp{CHEFS2020,Wright2021}), so a globally
distributed MELODI fleet would both close its own strong-wind calibration
gap and supply the globally representative in situ winds that satellite
calibration lacks. Wide, sustained deployment is therefore our central
practical recommendation.

The remaining avenues for improvement are incremental. The retrieval
already operates near the scatterometer accuracy floor
(section~\ref{sec:accfloor}), so the clearest gain would come from a
better validation reference (synchronous, collocated in situ anemometry)
rather than from added model complexity. Foreseeable extensions would
make explicit the variables now absorbed into the empirical fit. Candidates are atmospheric
stability, although a representative air-temperature measurement from so
small a platform is challenging (a COARE-type bulk-flux treatment would
then apply; \citealp{Fairall2003}); barometric pressure, valuable in its
own right for assimilation and entering the stress-equivalent wind
through air density \citep{deKloe2017}; wave--current interaction (the
GNSS velocity is already recorded); growth/decay hysteresis of the
equilibrium level \citep{Young1999,Donelan1985}; and explicit
spectral-shape descriptors in the analytical law. Each is a
self-contained refinement rather than a redefinition of the method.

\section{Summary and conclusions}

The open ocean still lacks systematic in situ wind observations, and the
calibration and validation of satellite scatterometers, radiometers, and
altimeters---and of the atmospheric and wave models they feed---depends
on the very collocated surface measurements that are absent over most of
the world ocean. This paper has shown that a compact, freely drifting
GNSS/IMU wave buoy---MELODI---can supply the missing wind vector at
accuracies useful for satellite calibration and validation, and at a
small fraction of the cost of moored platforms.

The wind speed retrieval (WITR) inverts the full shape of the measured
acceleration spectrum through Tikhonov-regularized regression rather than
reading a single equilibrium-range level, and wind direction is recovered,
with no fitted reference-wind regression, from IMU-derived directional
wave moments in the wind-sea band. Cross-validated by leave-one-buoy-out against Level-2 collocations
from four scatterometer products (ASCAT MetOp-B/C, HY-2B/C), WITR attains a
wind-speed RMSD of 0.91~m~s$^{-1}$, approximately half the
1.80~m~s$^{-1}$ obtained with the traditional single-band Toba inversion;
the IMU-derived direction reaches a mean absolute angular difference (MAAD) of
9.4$^{\circ}$ against ASCAT L2. The combined wind-vector difference is
competitive with that reported for moored anemometer buoys against the
same scatterometers. These figures are constrained over the
well-populated 4--14~m~s$^{-1}$ range; stronger winds are an
extrapolation, flagged as reduced confidence in the product.

A central result is that the data-driven model can be distilled, via
symbolic regression, into compact analytical equations that approach
WITR's performance while remaining interpretable and inexpensive enough
for onboard evaluation. These expressions expose physically interpretable
structure---wave-age modulation, smooth-to-rough transition behaviour,
and an empirical high-wind correction---linking the retrieval back to
equilibrium-range wind--wave theory.

In the product, WITR is the archived reference wind-speed retrieval; the
analytical equations are its deployable closed-form counterparts---the
extended spectral law [Eq.~(\ref{eq:pysr_c15})] when only the acceleration
spectrum is available, and the reduced drag law
[Eq.~(\ref{eq:reduced_drag})] when the IMU motion inputs are also
recorded.

Wave-sensing drifters are already an established source of the open-ocean wind
vector; these results show that a smaller platform can deliver both
components---including a markedly more reliable direction at low
wind---at an accuracy comparable to published scatterometer--buoy
differences, making it useful for satellite calibration and validation
and for data assimilation. The retrieval already operates close
to the accuracy floor set by the scatterometer reference, so the clearest
path to sharpening it further is not added model complexity but a more
precise validation reference---high-precision in situ anemometer
measurements taken synchronously and collocated with the drifter. Because
each MELODI is small,
low-cost, and already returns data reliably over multi-month deployments,
the natural next step is to scale the approach to a larger fleet,
building the density of open-ocean wind vectors that routine calibration,
validation, and assimilation require---observations that are still
lacking over most of the world ocean.

\clearpage
\acknowledgments

The author thanks the two anonymous reviewers, whose thorough and
constructive comments led to substantial improvements of this work.
This work would not have been possible without the involvement and help of
many people and organizations. The author is grateful to Lucas Charron
(eOdyn) for the design, development, manufacturing, and testing of the
MELODI buoys, and to the wider eOdyn team for their help with the MELODI
project and for their administrative and technical support. The author
thanks Bertrand Chapron for valuable discussions and remarks, and Fabrice
Collard (OceanDataLab) and Craig Donlon (ESA) for their help and support
during the ESA Ocean Training Course 2025 (OTC25) deployment campaign. The
OTC25 buoys were funded by CNES (Contract No.~5700012680); the EXPLOI buoy
was funded through the Exp\'{e}dition Plastique Oc\'{e}an Indien project
(COI, AFD, and FFEM), with the IRD as scientific partner.
This work was partially supported by the AI4COPSEC project under the
Horizon Europe programme.

\paragraph*{Competing interests.}
The author is employed by eOdyn, the company that develops and operates the
MELODI drifter evaluated in this study; the Sofar Spotter used as a
comparison platform is a commercial product of a third party. eOdyn had no
role in the design of the validation references (scatterometer and ERA5
products are produced by independent agencies) or in the leave-one-buoy-out
evaluation protocol. The author declares no other competing interests.

\datastatement

The MELODI data from the OTC25 deployment campaign are publicly available
from the SEANOE repository at \url{https://doi.org/10.17882/117337}. The
trained WITR model---including the linear TR stage, the gated GBM residual
correction, the feature-scaling parameters, the input-feature definitions,
the full catalogue of the 52 candidate features (definitions and ranked
importance) evaluated during model development, and example inference
scripts---is publicly archived on Zenodo at
\url{https://doi.org/10.5281/zenodo.22209123}. The reference datasets are
public: ERA5 reanalysis is distributed by the Copernicus Climate Data Store
(\url{https://cds.climate.copernicus.eu}); the ASCAT (MetOp-B/C) and HY-2B/C
Level-2 scatterometer winds, in the KNMI standard format, are produced by
EUMETSAT and the National Satellite Ocean Application Service
(NSOAS).

\bibliographystyle{ametsocV6}
\bibliography{references_v2}

\end{document}

%% file: tables/v14_coeffs_v3.tex
\begin{table}[t]
\caption{Linear (TR) stage of the WITR model: per-feature standardization constants and Ridge coefficients. Bands in Hz; $\overline{S}_{\mathrm{acc}}$ is the band-mean acceleration-spectral density; the 0.60--0.80 entry is the band-median high-frequency level. The GBM correction stage is provided with the archived model (see Data Availability).}
\label{tab:v14_coeffs}
\centering
\footnotesize
\setlength{\tabcolsep}{3.5pt}
\begin{tabular}{lrrr}
\hline\hline
        & Scaler & Scaler & Std.\ Ridge \\
Feature & mean   & std    & coeff. \\
\hline
$\overline{S}_{\mathrm{acc}}$ 0.12--0.18 & 1.2494 & 0.9793 & 0.4894 \\
$\overline{S}_{\mathrm{acc}}$ 0.18--0.25 & 1.6536 & 0.9755 & 0.9775 \\
$\overline{S}_{\mathrm{acc}}$ 0.25--0.35 & 1.7847 & 0.7992 & 1.0767 \\
$\overline{S}_{\mathrm{acc}}$ 0.35--0.50 & 1.6600 & 0.5625 & 0.7048 \\
$\overline{S}_{\mathrm{acc}}$ 0.50--0.70 & 1.3436 & 0.3166 & $-$0.2328 \\
Noise floor 0.60--0.80 & 0.4478 & 0.5791 & $-$0.1388 \\
Slope 0.50--1.00 & $-$1.1007 & 0.4704 & $-$0.3297 \\
Slope 0.25--0.50 & $-$0.0201 & 0.8105 & $-$0.0647 \\
$f_{25}$ 25th-pctl.\ freq. & 0.3147 & 0.0576 & 0.7221 \\
\hline
Intercept (m\,s$^{-1}$) & & & 7.8166 \\
\hline\hline
\end{tabular}
\end{table}